\PassOptionsToPackage{expansion=false}{microtype}
\documentclass[sigconf, nonacm]{acmart}

\usepackage{pvldb}

\renewcommand\vldbdoi{XX.XX/XXX.XX}
\renewcommand\vldbpages{XXX-XXX}
\renewcommand\vldbavailabilityurl{}

\newcommand{\opFanoutReshapeJoin}{\textsc{Fanout-Reshape Join}}
\newcommand{\opKeyAlignedSIMDJoin}{\textsc{Key-Aligned SIMD Join}}
\newcommand{\opVPUIndexLookupJoin}{\textsc{VPU Index Lookup Join}}
\newcommand{\opSparseGatherJoin}{\textsc{SparseCore Gather Join}}
\newcommand{\opSortedHashLookupJoin}{\textsc{Sorted-Hash Lookup Join}}
\newcommand{\opOneHotMatMulAgg}{\textsc{One-Hot MatMul Aggregation}}
\newcommand{\opVMEMBlockHistogram}{\textsc{VMEM Block Histogram}}
\newcommand{\opSparseSegmentReduction}{\textsc{Segment Reduction}}
\newcommand{\opFanoutReshapeAgg}{\textsc{Fanout-Reshape Aggregation}}
\newcommand{\opSortedRunAgg}{\textsc{Sorted-Run Aggregation}}
\newcommand{\opRadixPackedAgg}{\textsc{Radix-Packed Aggregation}}
\newcommand{\opMXUOneHotDistinct}{\textsc{MXU One-Hot Distinct}}
\newcommand{\opVMEMBitonicDistinct}{\textsc{VMEM Bitonic Distinct}}
\newcommand{\opCompactSortDistinct}{\textsc{Compact-Then-Sort Distinct}}
\newcommand{\opDirectSortDistinct}{\textsc{Direct-Sort Distinct}}
\newcommand{\opInKernelTopK}{\textsc{In-Kernel Top-K}}
\newcommand{\opGlobalSort}{\textsc{Global Sort}}
\newcommand{\opHostHybridSort}{\textsc{Host-Hybrid Sort}}

\definecolor{ntaFigureBlue}{HTML}{0072B2}
\definecolor{ntaFigureBlueLight}{HTML}{DFEDF5}
\definecolor{ntaFigureInk}{HTML}{4B5563}
\definecolor{ntaFigureGray}{HTML}{9CA3AF}
\definecolor{ntaFigureHairline}{HTML}{BEC2CA}
\newcommand{\para}[1]{\noindent\textbf{#1.}}
\usepackage{multirow}
\usepackage{enumitem}
\usepackage{tikz}
\usetikzlibrary{arrows.meta,calc,positioning,shapes.geometric}

\begin{document}
\title{The Case for TPU-Accelerated Analytical Query Processing}

\author{Jaehyun Ha, Yeounoh Chung\texorpdfstring{\textsuperscript{*}}{*}, Helena Caminal, Aditya K Kamath,\\Fatma \"Ozcan, Arvind Krishnamurthy, Henry M. Levy}
\thanks{\textsuperscript{*}Corresponding author.}
\affiliation{%
  \institution{SystemsResearch@Google}
  \city{Mountain View}
  \country{USA}
}
\email{{jaehyunha, yeounoh, hcaminal, akkamath, fozcan, arvindkrish, hanklevy}@google.com}

\begin{abstract}
Large-scale analytical query processing requires computing substrates capable of immense distributed scale-out. Tensor Processing Units (TPUs) perfectly embody this extreme-scale architecture, offering massive computational density and highly scalable optical pod-level interconnects. Despite these advantages, TPUs have so far remained largely unexplored for analytical database workloads.
In this paper, we present the first comprehensive case for TPU-accelerated analytical query processing by running all the TPC-H queries entirely on the TPUs. We introduce Native TPU Accelerator (NTA) for analytical query processing, which provides hardware-conscious algorithm design and optimizations in JAX/XLA. By exposing the TPU's structural mechanics, we demonstrate how dynamic, memory-bound database operations can be rewritten into deterministic JAX programs. We engineer a suite of novel execution strategies that leverage both the TPU's TensorCore and SparseCore to trade sparse, unpredictable memory accesses for dense, hardware-aligned ones. At the core of NTA, a compiler uses catalog statistics to assign guarded static capacities and a hardware-calibrated local cost model to select a TPU-native implementation for each relational operator before JIT compilation.

We evaluate our system using the TPC-H benchmark. Taking full advantage of TPU slices and JAX distributed computing, we seamlessly scale our execution environment to run TPC-H benchmark up to Scale Factor 1000 (SF1000). Evaluated against a rigorous CPU ground truth (DuckDB), our native TPU accelerator achieves \textit{state-of-the-art} performance, achieving a $7.0\times$ single-device speedup over DuckDB at SF100 and a $68.6\times$ speedup with 16 TPUs at SF1000 (1TB).
The results establish TPUs not merely as dense neural compute hardware, but as a highly competitive, massively scalable substrate for the next generation of analytical query processing.
\end{abstract}


\maketitle

\vldbtopmatter

\section{Introduction}

Hardware-accelerated query processing is gaining significant momentum across both academic and commercial analytics systems~\cite{TQP, CoddSpeed,yong2025evaluating,cao2023gpu,mostak2016mapd}, which require computing substrates capable of immense, distributed scale-out. 
Tensor Processing Units (TPUs) perfectly embody this capability: powering massively large-scale search and AI/ML workloads, they offer extreme computational density and highly scalable optical pod-level\footnote{A TPU Pod is a cluster of Google Tensor Processing Units (TPUs) connected by a high-speed network to act as a single, powerful AI supercomputer.} interconnects~\cite{TPUv4}. Despite these massive structural advantages, TPUs remain largely unexplored for analytical database systems. This paper fundamentally explores running relational queries on TPU architectures, revealing precisely what it takes to map analytical queries to what is conventionally assumed to be a static execution paradigm built explicitly without hardware-managed caches or dynamic execution.

TPU and especially its main compute units, TensorCores, require a fixed and deterministic execution environment where memory access and shapes must be compiled and pre-determined. In contrast, relational operators rely on data-dependent dynamism~\cite{StableHLO}, which is hard to predict.
The number of rows produced by a filter (\textit{shape dynamism}) and the memory locations where matching tuples reside in a join (\textit{address dynamism}) are strictly runtime properties determined by the data. The TPU TensorCore, however, enforces completely static evaluation: every instruction, buffer size, and DMA transfer must be predictably scheduled at query compile time. Naively mapping relational operations onto such a static substrate creates immense \textit{sparsity} out of forced worst-case padding, and \textit{HBM pipeline stalls} resulting from irregular gather/scatter\footnote{\textit{Gather} refers to indirect, non-contiguous reads (e.g., retrieving join matches via index vectors); \textit{scatter} refers to indirect writes or in-place accumulations at arbitrary addresses (e.g., updating hash buckets or histogram bins).}. Traditional CPUs and GPUs mask these erratic access patterns dynamically using large, hardware-managed caches and massive multi-threading architectures~\cite{Tesla}.
Fortunately, modern TPUs added SparseCore units to accelerate sparse, non-uniform memory accesses that can run in parallel with TensorCore computations.

In this work, we argue that TPUs can be used as efficient processors for analytical workloads
by fundamentally neutralizing dynamic memory accesses, using a new TPU hardware co-processor, SparseCore~\cite{jaxdocs}. We introduce Native TPU Accelerator (NTA), a tensor-based execution engine built entirely in JAX and XLA~\cite{jax2018github}~\footnote{JAX is an ML framework that represents array operations as tensor computation graphs; XLA (Accelerated Linear Algebra) is an ML compiler that lowers these computation graphs into statically scheduled, fused executables on HW accelerators.}. 
Rather than treating the TPU as a black-box HW accelerator, NTA co-designs relational operators directly around its memory and compute hierarchy. Our core insight is that dynamic, memory-bound database operations can be meticulously rewritten into predictable, compute-bound arithmetic. For instance, \opSortedHashLookupJoin{} avoids sparse, padded hash tables, using unique-build equi-joins by sorting hashed composite keys and validating each retrieved candidate against the original packed key.

Providing highly optimized kernels is insufficient unless a dynamic relational plan can be converted into a well-orchestrated TPU program.
To achieve this, NTA introduces an end-to-end query compilation and execution pipeline (Section~\ref{sec:system-overview}).
Given an incoming logical plan, NTA uses catalog statistics to assign guarded capacities (bounded static shapes) to intermediate results, derives existing logical and physical properties, and selects a cost-optimal execution target (TensorCore vs. SparseCore) kernel implementation for each operator. NTA then stitches individual physical operators into a physical plan, which is then lowered (program transformation)  into a statically shaped JAX program with Pallas kernels~\footnote{Pallas kernel is a custom, low-level function in Python that runs directly on hardware accelerators (e.g., TPUs) within the JAX framework~\cite{jaxdocs}.}. Finally the JAX program is compiled and further optimized using the XLA compiler.

To evaluate NTA at large scales (up to SF1000), we leverage JAX distributed (XLA SPMD~\cite{xu2021gspmd}) computing over  TPU v5p~\cite{googlecloud_tpuv5p} slices. 
For a rigorous baseline, we compare our performance against DuckDB, a state-of-the-art SIMD-vectorized in-memory CPU database engine~\cite{CoddSpeed,GPUDB,TQP}. 
NTA achieves state-of-the-art performance, outperforming DuckDB on all 22 TPC-H queries and achieving geometric-mean speedups of $7.0\times$ at SF100 (100GB) and $68.6\times$ at SF1000 (1TB). When measured by total query-set runtime, the corresponding total-runtime speedups are $5.7\times$ and $50.1\times$ over the CPU baseline.

In summary, this paper makes the following contributions:
\begin{itemize}
    \item \textbf{The case for TPU-Accelerated Analytical Query Processing}: We present the first deep architectural analysis of mapping relational algebra onto TPU semantics, explicitly identifying core anti-patterns (shape dynamism, sparsity, gather/scatter stalls) and standardizing their mitigation.
    \item \textbf{Native TPU Kernels via Pallas \& SparseCore:} We introduce a suite of highly optimized, TPU-native execution strategies 
    , designed to address gather/scatter bottlenecks and HBM anti-patterns (round-trips and fusion breaks).
    \item \textbf{Compilation and Local Physical Planning:} Given a fixed logical plan, NTA bounds intermediate cardinalities, filters physical implementations by applicability and resource constraints, and selects among them using TPU v5p-calibrated operator costs before JIT compilation.
    \item \textbf{Comprehensive Evaluation at Extreme Scale:} We demonstrate state-of-the-art analytical performance on TPC-H, utilizing JAX distributed computing to scale smoothly to SF1000 across orchestrated TPU slices.
\end{itemize}

The remainder of this paper is organized as follows: Section~\ref{sec:background} provides the architectural background on TPU and highlights our motivating micro-benchmarks. Section 3 details our native TPU-accelerated query processing algorithms. Section 4 describes our compiler and hardware-conscious intelligent kernel selection. Section 5 details extreme scale orchestration. Section 6 presents our extensive experimental evaluation. Finally, Sections 7 and 8 offer discussion and conclusions.

\section{Background} \label{sec:background}

\subsection{TPU Architecture}
To understand the difficulty of porting database workloads to TPUs, one must examine the execution model. A modern TPU chip hosts \textit{TensorCores}, each combining Matrix Multiply Units (MXUs)---systolic arrays for dense matrix math---with a Vector Processing Unit (VPU) for element-wise operations. Each TensorCore owns a \textit{VMEM}, a software-managed scratchpad serving as its fast memory. Unlike a hardware-managed CPU or GPU cache, VMEM is filled and drained exclusively by explicit DMA transfers, each statically described at program/binary compile time as a burst with a base address, stride, and count. 
Crucially, sizes and exact memory addresses need to be known at compile time to generate the memory instructions and DMA descriptors to move the right data to the vector lanes of the TensorCore.

To appreciate the potential of the TPU for analytical query processing, one must understand how fundamentally its design differs from traditional hardware. Rather than relying on hardware-managed caches, branch predictors, or fine-grained thread-level parallelism to intuitively hide the latency of irregular memory accesses, the TPU operates on a philosophy of deterministic simplicity. Memory is explicitly managed via software, and execution proceeds via a statically scheduled Very Long Instruction Word (VLIW) pipeline. 
This extreme architectural minimalism makes TPUs vastly more efficient at predictable, dense tensor mathematics, but creates a challenging (but tractable) environment with the data-dependent dynamism inherent in relational algebra.

\begin{figure}[t!]
    \centering
    \includegraphics[width=0.48\textwidth]{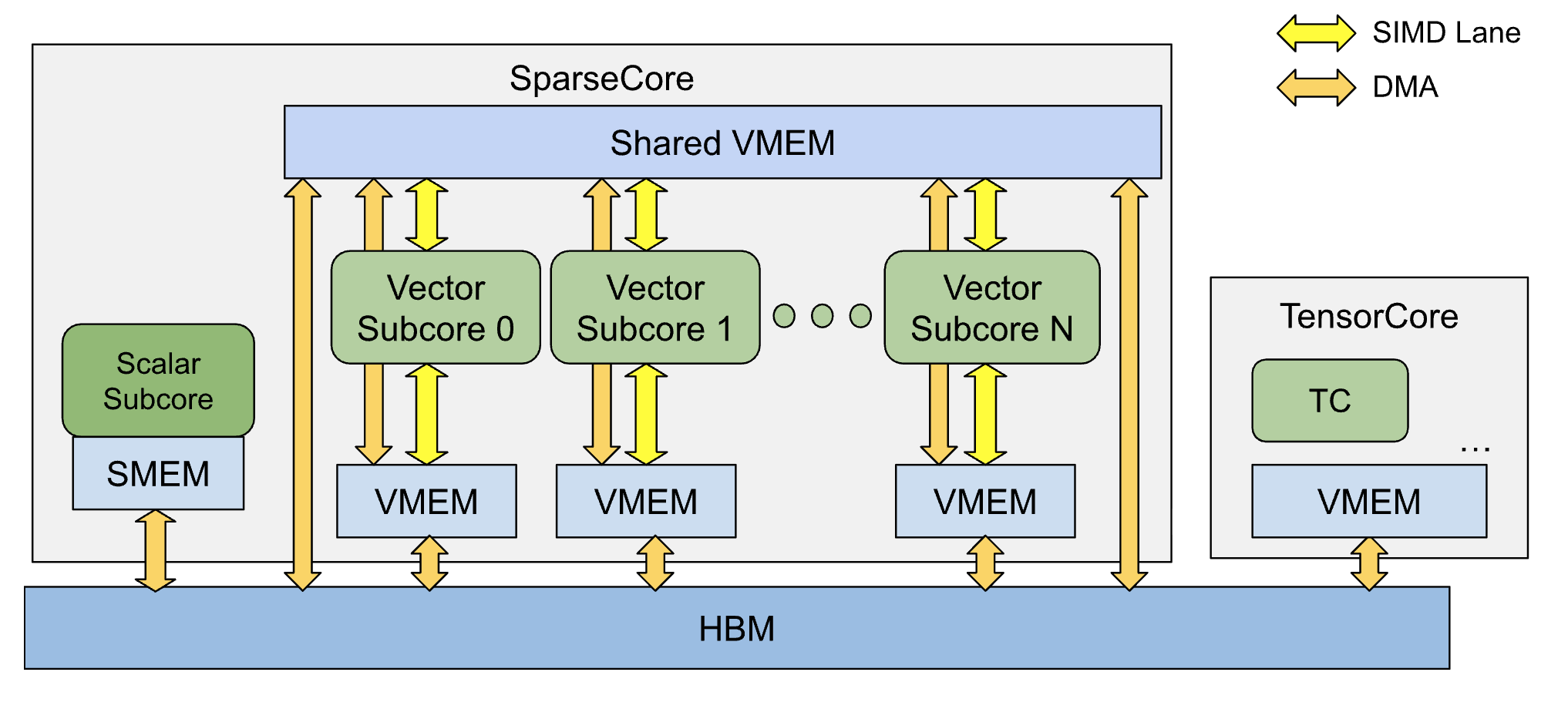}
    \caption{TPU TensorCore and SparseCore HW overview and memory hierarchy \cite{jaxdocs}. } \vspace{-0.3cm}
    \label{fig:sparsecore}
\end{figure}

One key component of the TPU is the \textit{SparseCore}, recently added as an ally to the TensorCore. SparseCores traverse irregular index streams, performing gathers and scatters from HBM, on behalf of TensorCores or as their own compute engines.
However, SparseCore and TensorCore programs inherently exist separately and transfers between them require explicit DMA staging and synchronization via CMEM (common memory).
These architectural tradeoffs share a single premise: a careful mapping of operators to the right cores in the TPU is necessary.
Table~\ref{tab:dynamism} contrasts these choices with contemporary hardware for SQL query processing.

\begin{table*}[t]
  \caption{The same decisions, made at different times.}
  \label{tab:dynamism}
  \begin{tabular}{llll}
    \toprule
    Decision & CPU: at runtime & GPU: at runtime & TPU: at compile time\\
    \midrule
    \small\textbf{Where to read/write} & Any memory dynamically & Any register dynamically & Fixed in the program \\
    \small\textbf{How to hide latency} & Caches \& Pre-fetching & Massive multithreading & Statically scheduled DMA \\
    \small\textbf{What to keep on chip} & Multi-level cache (L1-L3) & Hardware caches (L1/L2) & Compiler-managed VMEM \\
    \small\textbf{How much to allocate} & Dynamic virtual memory & Dynamic scaling & Static worst-case bounds OR Use SC \\
  \bottomrule
\end{tabular}
\end{table*}

\subsection{The Software Stack and Tensor Query Processing} \label{sec:background:sw}
The software stack enforces this compile-time discipline end to end. JAX traces Python programs into a functional intermediate representation where every tensor has a fixed shape---unknown sizes must be padded to a static bound. XLA lowers this representation to TPU code, allocating every buffer and scheduling every DMA transfer ahead of execution. Beneath XLA, Pallas allows programmers to write custom TPU kernels with explicit control over blocking and the HBM-VMEM pipeline, enabling advanced query optimizations (Section~\ref{sec:operators}).

XLA’s most critical optimization is operator \textit{fusion}. XLA fuses chains of operations into a single loop, streaming intermediate results through VMEM without touching HBM, sustaining high compute utilization. However, fusion relies on \textit{regular} memory accesses expressible as static DMA bursts. When an access depends on data (e.g., a hash join gather reading $A[idx[i]]$), no DMA burst can describe it, snapping the computational chain. The gather may touch any element, forcing the entire table to reside in HBM. We call this a \textit{fusion break}---the tensor-program analogue of a pipeline breaker in query engines, but far stricter. On a TPU, merely executing a single data-dependent gather breaks the computation pipeline, forcing massive HBM roundtrips.

Within this static ecosystem, our starting point leverages the Tensor Query Processing strategy~\cite{TQP,CoddSpeed}, which compiles relational queries into columnar tensor programs. GPUs excel at this: a selection equates to a boolean mask, and a join relies on index computations followed by gathers. GPUs mask the latency of dynamic gathers using hardware-managed caches and SIMT multithreading optimizations. These mappings were designed for hardware that resolves dynamism at runtime; however, when ported directly to a TPU, performance collapses due to the dynamism (Section~\ref{sec:dynamism}).

\subsection{The Gather/Scatter Bottleneck}
In relational processing, joins and aggregations map to pipeline-breaking gathers and scatters, associated with TPU HBM roundtrips (\textit{Anti-Pattern}). To quantify the cost of this \textit{Anti-Pattern}, we established a microbenchmark framework on the Google TPU v5p~\footnote{TPU v5p has 2 TensorCores and 4 SparseCores, 95GB HBM, see \url{https://docs.cloud.google.com/tpu/docs/v5p}}. We evaluated a 4-Byte Integer Gather (a canonical analytical lookup) mapping a 60-million-row outer relation against a 15-million-row inner relation. 
Table~\ref{tab:gather} maps different memory access patterns to potential gather bottlenecks (Queue) and resolutions. While the hardware pipeline is highly efficient under high-locality/low-sparsity, the \textit{Clustered High Sparsity} access pattern (e.g., highly selective SQL queries) results in a severe performance bottleneck.

\begin{table}[!th]
  \caption{Gather Bottleneck Analysis (60M Probe $\rightarrow$ 15M Build)}
  \label{tab:gather}
  \resizebox{\columnwidth}{!}{%
  \begin{tabular}{p{0.24\linewidth} p{0.24\linewidth} p{0.4\linewidth}}
    \toprule
    Access Pattern & Queue & Resolution \\
    \midrule
    \small
    \textbf{High Locality \& Low Sparsity} & Low (Balanced DMA requests) &  Subcores operate near peak HBM BW bounds. \\
    \small
    \textbf{Random \& High Sparsity} &  Moderate (Uniform queueing) & Subcores idle due to unpredictable random memory accesses ($\tau_{in}$). \\
    \small
    \textbf{Clustered \& High Sparsity} & \textbf{Extreme (Funnel Failure)} & A single subcore absorbs 100\% of memory bank contention ($\tau_{ac}$). \\
  \bottomrule
\end{tabular}
}
\end{table}

\begin{figure}[!th]
  \centering
  \includegraphics[width=\linewidth]{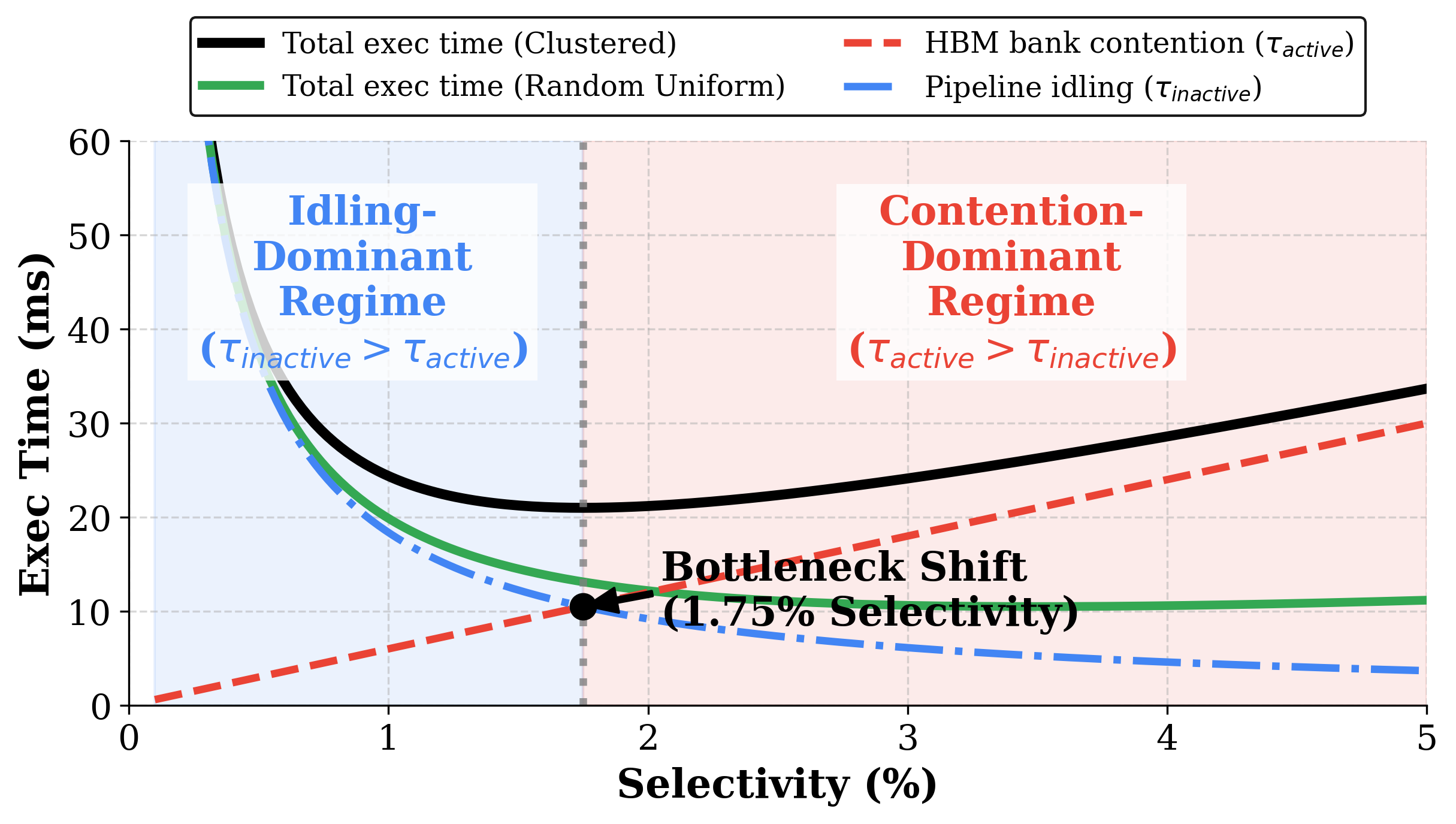}
  \caption{Gather execution time (60M $\rightarrow$ 15M) under varying selectivity. The intersection demonstrates where hardware bank contention ($\tau_{active}$) overtakes pipeline idling cost ($\tau_{inactive}$) as the dominant bottleneck at 1.75\% selectivity.}
  \label{fig:microbenchmark}
\end{figure}

As illustrated in Figure~\ref{fig:microbenchmark}, decomposing the gather execution time across varying selectivities reveals a critical inflection point.
At query selectivities dropping to 1.75\%, HBM bank contention $\tau_{active}$ overtakes pipeline idling ($\tau_{inactive}$) as the primary performance bottleneck. Under clustered and highly selective access patterns, the most heavily loaded SparseCore subcore lane suffers from a ``funnel failure,'' absorbing queued DMA access that serializes memory access.

\subsection{Data-Dependent Dynamism on TPUs} \label{sec:dynamism}
The gather/scatter bottlenecks highlight a fundamental architectural tension: 
relational query processing is inherently data-dependent, whereas the TPU's compile-time optimization stack (XLA) relies on static shapes and deterministic execution graphs for performance optimization. This section defines the resulting performance gaps and presents the core design principles that guide our Native TPU Accelerator (NTA).

\para{A Taxonomy of Dynamism} Data-dependent dynamism manifests in two distinct forms. \emph{Shape dynamism} arises when the output cardinality of an operator depends on its input data. This imposes the cost of \emph{sparsity}: every dynamically sized tensor must be declared at a worst-case bound, and every padded element is computed. \emph{Address dynamism} arises when \emph{where each row belongs} in the output is decided by runtime key values (e.g., gathers and scatters). This imposes the cost of \emph{stalls}: every data-dependent access breaks operator fusion, forces its intermediate to be materialized in HBM, and stalls the TensorCore while it waits for the HBM round trip.
Moreover, these two forms of dynamism compound each other. When address dynamism forces an intermediate tensor to materialize in HBM, it must do so at the worst-case size (for the output tensor) dictated by shape dynamism.

\para{The Compound Cost of Sparsity and Stalls} Sparsity taxes every unit on the chip. When sweeping the fraction of padding, the execution time of operations on both the TensorCore and SparseCore stays \emph{flat} on the TPU, as padded rows consume identical cycles to valid rows. 
Address dynamism introduces even a deeper cost; inserting a single data-dependent gather into a fused chain of element-wise operators forces the XLA compiler to inevitably split the chain. This is because the compiler cannot fuse element-wise operations across a dynamic look-up or gather.
When the fusion breaks, the TPU is forced to materialize intermediate results to HBM, causing execution stalls on the TensorCore.
While offloading the gather to the SparseCore mitigates the inefficiency of using TensorCore systolic arrays for scalar lookups, the mathematical operations chain (fusion) still breaks causing the TensorCore units to stall while waiting for the HBM roundtrip.

Crucially, these costs multiply. Worst-case bounds compose across operators, generating massive padding factors in deep pipelines. And every materialization heavily inflates the HBM stall, as it is written at the worst-case, padded size. A single operator may survive dynamism; a query plan does not.

\para{System Design Implications} These constraints translate into three non-negotiable requirements for a TPU accelerator. \textbf{R1:} dynamic shapes must be bounded \emph{tightly}. \textbf{R2:} data-dependent memory accesses must be mathematically translated into predictable tensor arithmetic (e.g., boolean masking) to sustain pipelined execution and avoid fusion breaks. \textbf{R3:} unavoidable dynamic accesses must execute via SparseCore and its Scatter-Gather Engine (SGE) rather than relying on the TensorCore path.
\vspace{-0.3cm}

\section{NTA System Overview}
\label{sec:system-overview}

\begin{figure*}[!th]
  \centering
  \includegraphics[width=\textwidth]{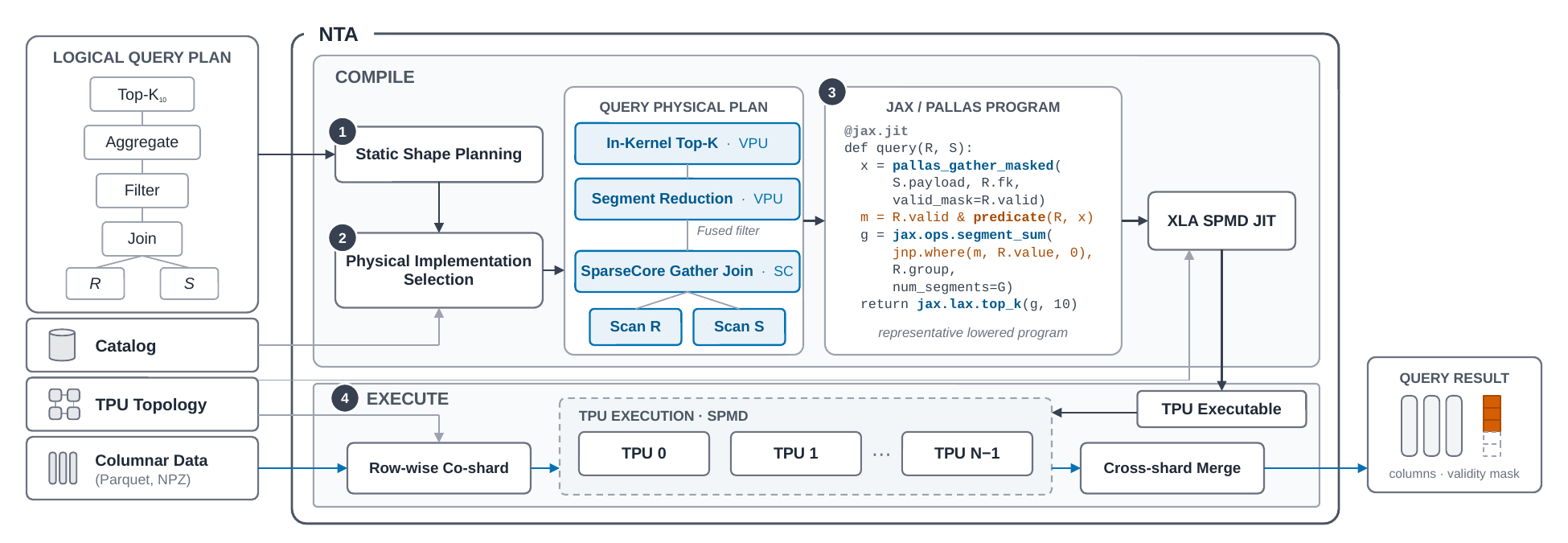}
  \vspace{-8mm}
  \caption{
  \textbf{NTA Architecture and End-to-End Query Lifecycle.} (1) Static shape planning assigns guarded buffer bounds; (2) Physical implementation selection maps operators to VPU, MXU, or SparseCore; (3) Plan lowering emits a JAX/Pallas program compiled by XLA SPMD; (4) Distributed execution evaluates queries over co-partitioned shards across the TPU slice.}
  \Description{A logical query plan and catalog statistics feed intermediate-cardinality bounding and physical operator selection. The resulting physical query plan is lowered to JAX/Pallas and compiled by XLA SPMD for the assigned TPU topology. Input relations are co-partitioned across TPU devices, which execute the compiled query over their local partitions before partial outputs are combined into the final result relation.}
  \label{fig:system-overview}
\end{figure*}


Modern TPUs offer distinct architectural advantages for massive tensor computation throughput, dedicated hardware gather-scatter acceleration (SparseCore), programmable on-chip scratchpad memories, and ultra-fast ICI inter-chip networks. 
In order to unlock theses hardware capabilities for analytical query processing, we present \textbf{Native TPU Accelerator (NTA)}. Rather than treating the TPU as a black-box coprocessor, NTA co-designs relational operators directly around its memory and compute hierarchies. NTA selects highly optimized physical Pallas kernel implementations, further optimizes and distributes its execution over available TPU topology and across different TPU computing units (e.g., TensorCore, SparseCore).
Figure~\ref{fig:system-overview} illustrates the compilation and distributed execution workflow of NTA.

\begin{enumerate} [leftmargin=4mm]
    \item \textbf{Plan Ingestion and Shape Planning (Section~\ref{sec:compiler:shape}):} NTA ingests logical query plans and uses catalog statistics to compute compile-time cardinality bounds ($\mathcal{B}_o$) for every intermediate relation, sizing static tensor buffers with calibrated safety margins.
    \item \textbf{Physical Operator Selection (Section~\ref{sec:operators}, \ref{sec:compiler:decision}):} Using compiled bounds, key distributions, and target TPU hardware cost profiles, NTA selects the optimal physical operator from its TPU-native catalog (e.g., Sorted-Hash Lookup Join vs. SparseCore Gather Join).
    \item \textbf{Lowering to JAX and Pallas (Section~\ref{sec:compiler:lowering}):} Non-materializing operators (filters, projections) are fused into adjacent kernels. The physical plan is lowered to a statically shaped JAX program with embedded Pallas custom kernels for specialized memory-movement routines.
    \item \textbf{XLA SPMD Compilation and Scale-Out  (Section~\ref{sec:orchestration}):} The lowered program is compiled via XLA SPMD into a single device-mesh executable. During execution, relations are co-partitioned across TPU devices, each device processes local partitions via 32-bit local tuple identifiers, and partial results are merged into the final relation.
\end{enumerate}

\section{TPU-Native Relational Operators}
\label{sec:operators}

NTA represents each intermediate relation as statically shaped column tensors with a validity mask and compiles relational operators into JAX or Pallas programs. Selections and projections are fused into the next materializing operator, whereas joins, aggregations, duplicate elimination, and sorting may each admit multiple TPU-native implementations. Table~\ref{tab:rewrite_map} summarizes this physical operator catalog, including the input data property required by each implementation and its key execution primitives. The following subsections explain how each implementation preserves relational semantics and why its listed property makes it applicable. Section~\ref{sec:compiler} then describes how NTA combines these properties with planned cardinality bounds and calibrated crossover rules (e.g., ``small'' domain) to select among the applicable implementations.

\begin{table*}[t]
  \caption{NTA physical operator catalog.}
  \vspace{-2mm}
  \label{tab:rewrite_map}
  \resizebox{\textwidth}{!}{
  \begin{tabular}{@{}lllll@{}}
    \toprule
    \textbf{Logical Op.} & \textbf{Physical Operator} & \textbf{Required Property} & \textbf{Primary Unit} & \textbf{Key Primitive(s)} \\
    \midrule
    \multirow{5}{*}{\textbf{Join}}
      & \opKeyAlignedSIMDJoin      & 1:1 join; identical row-aligned keys & VPU       & \texttt{jnp.bitwise\_and} \\
      & \opFanoutReshapeJoin       & Clustered PK--FK; bounded fanout $k$ & VPU        & \texttt{jnp.reshape}, \texttt{jnp.repeat} \\
      & \opVPUIndexLookupJoin      & PK--FK; existing lookup index & VPU          & \texttt{jnp.take} \\
      & \opSparseGatherJoin        & PK--FK; existing lookup index & SparseCore                & \texttt{plsc.Indices}, \texttt{pltpu.emit\_pipeline}, \texttt{pltpu.sync\_copy} \\
      & \opSortedHashLookupJoin    & Unique-build equijoin & VPU                          & \texttt{integer\_hash}, \texttt{jnp.argsort}, \texttt{jnp.searchsorted}, \texttt{jnp.equal} \\
    \midrule
    \multirow{6}{*}{\textbf{Aggregation}}
      & \opOneHotMatMulAgg         & Small dense group domain & MXU                         & \texttt{jax.nn.one\_hot}, \texttt{jax.lax.dot\_general} \\
      & \opVMEMBlockHistogram      & Histogram working set fits VMEM & VPU                    & \texttt{pl.pallas\_call}, \texttt{pl.atomic\_add} \\
      & \opRadixPackedAgg          & Multi-key; small packed domain & MXU                       & \texttt{jnp.ravel\_multi\_index}, \texttt{jax.lax.dot\_general} \\
      & \opSortedRunAgg            & Input ordered by group key & VPU                           & \texttt{jnp.diff}, \texttt{jnp.cumsum} \\
      & \opFanoutReshapeAgg        & Clustered groups; bounded fanout $k$ & VPU                  & \texttt{jnp.reshape}, \texttt{jnp.sum} \\
      & \opSparseSegmentReduction  & Large or sparse group domain & VPU, SparseCore                       & \texttt{jax.ops.segment\_sum} \\
    \midrule
    \multirow{4}{*}{\textbf{Distinct}} 
      & \opMXUOneHotDistinct       & Small dense value domain & MXU                          & \texttt{jax.nn.one\_hot}, \texttt{jax.lax.dot\_general} \\
      & \opVMEMBitonicDistinct     & Active working set fits VMEM & VPU                       & \texttt{jax.lax.sort} \\
      & \opCompactSortDistinct     & Few active rows after filtering & VPU                      & \texttt{jnp.nonzero}, \texttt{jax.lax.sort} \\
      & \opDirectSortDistinct      & General fallback & VPU                                     & \texttt{jnp.unique(size=...)} \\
    \midrule
    \multirow{3}{*}{\textbf{Sort / Top-K}}
      & \opInKernelTopK            & \texttt{ORDER BY ... LIMIT $K$} & VPU                    & \texttt{jax.lax.top\_k} \\
      & \opGlobalSort              & Full ordering on TPU & VPU                               & \texttt{jnp.argsort} \\
      & \opHostHybridSort          & Host candidate set $\le20$K rows & TPU + Host                & \texttt{jax.device\_get}, \texttt{np.argsort} \\
    \bottomrule
  \end{tabular}
  }
\end{table*}

\subsection{Joins} \label{sec:operators:joins}

NTA specializes join kernels to exploit physical properties of the input relations, including row alignment, bounded PK--FK fanout, and build-key uniqueness. The following paragraphs describe three implementation families: layout-specialized joins, indexed PK--FK joins, and sorted-hash joins over a unique build key. Indexed PK--FK joins assume a prebuilt lookup index, whereas sorted-hash joins construct their lookup structure on the fly when selected.

\para{Layout-specialized joins}
A clustered PK--FK layout stores at most $k$ child rows for each parent in fixed-width slots. \opFanoutReshapeJoin{} views the child columns as parent-by-$k$ tensors and broadcasts each parent payload across its
child slots. Unused slots remain invalid, preserving join semantics without an indexed lookup. \opKeyAlignedSIMDJoin{} applies to a 1:1 join whose inputs are already aligned on the join key: at every row position, the two inputs contain the same key. Having the same key domain is not sufficient because the relations may order their keys differently or contain different key multiplicities.

\para{Indexed joins}
A PK--FK lookup index maps each probe key to a build-side row identifier. \opVPUIndexLookupJoin{} gathers the corresponding payloads with \texttt{jnp.take}, whereas \opSparseGatherJoin{} sends the same row identifiers through \texttt{plsc.Indices} and a pipelined SparseCore transfer. Unmatched probes are directed to a safe sentinel position
and invalidated by the output validity mask. The multi-payload specialization reuses one row-identifier vector across multiple projected columns, amortizing the index lookup.

\para{Sorted-hash lookup}
For a unique-build equijoin without a prebuilt lookup index, \opSortedHashLookupJoin{} constructs a temporary lookup structure
during query execution. It packs and hashes the build keys, sorts the build rows by hash, and uses binary search to locate the matching hash range for each probe. It then scans a bounded number of candidates and compares the original keys before accepting a match, preventing hash collisions from producing false matches. An overflow guard detects a
collision range that exceeds the compiled scan bound. Build-key uniqueness guarantees at most one exact match per probe row.

\para{Join semantics}
All implementations expose a common output contract: a build-side row identifier and a match predicate for each probe row. Inner joins
invalidate unmatched probes, whereas left-outer joins retain the probe row and invalidate the unmatched build-side payloads. Semi and anti joins consume only the match predicate without materializing build-side payloads. A bounded-fanout layout extends the same contract to at most $k$ child slots per probe. The physical implementation therefore changes how matches are located without changing the logical join semantics.

\subsection{Aggregations} \label{sec:operators:agg}

Aggregation implementations differ in how they map grouping keys to aggregate state. NTA specializes this mapping to properties of
the grouping keys and input layout: a bounded dense key domain, an aggregate-state array that fits in VMEM, an existing group-key order,
or a clustered layout with bounded fanout. When none of these applies, NTA uses segment reduction as the fallback.

\para{Dense-domain aggregation}
When group identifiers form a bounded domain (group key range), \opOneHotMatMulAgg{} encodes them as a masked one-hot matrix and evaluates additive aggregate state with \texttt{jax.lax.dot\_general} on the MXU. For a composite grouping key, \opRadixPackedAgg{} first maps each key tuple to a dense integer
identifier with \texttt{jnp.ravel\_multi\_index}. This implementation applies when the product of the domains remains within the calibrated bound for the dense reduction.

\para{Histogram and segment aggregation}
When the aggregate state array fits in VMEM, \opVMEMBlockHistogram{} processes the input in tiles and updates the
group state on chip before writing the completed state to HBM. \opSparseSegmentReduction{} handles large or sparse group domains through the statically shaped reduction exposed by \texttt{jax.ops.segment\_sum}. Unlike the dense-domain implementations, these operators do not materialize an input-cardinality-by-domain-size one-hot matrix.

\para{Structure-aware aggregation}
\opSortedRunAgg{} applies when the input is already ordered by the grouping key. It identifies contiguous runs of equal keys and reduces each run into one group result. \opFanoutReshapeAgg{} applies when a
clustered layout stores at most $k$ input rows for each group in fixed-width slots; it reshapes the input into group-by-$k$ tensors and reduces along the slot dimension. These implementations use the existing order or layout rather than sorting or repartitioning the input.

\para{Aggregate state} 
NTA represents the SQL aggregates associated with each group as fixed-width state. \texttt{COUNT} and \texttt{SUM} use additive
accumulators, \texttt{AVG} maintains a sum--count pair, and \texttt{MIN} and \texttt{MAX} use masked extrema. Aggregates sharing the same grouping keys also share the group-to-state mapping. The product of the group-domain size and aggregate-state width determines
the state footprint and, in particular, whether the VMEM-resident histogram implementation is applicable.

\subsection{Distinct}
\label{sec:operators:distinct}

Distinct implementations differ in how they represent value membership. For a single integer-valued expression with a bounded
domain, NTA records the values present in the input directly. Otherwise, it sorts value tuples and retains one row from each
equality run. The sort-based implementations differ in how much inactive data they carry into the sort and whether the resulting working set fits the on-chip memory budget.

\para{Dense-domain distinct}
When the encoded value domain is within the calibrated one-hot limit, \opMXUOneHotDistinct{} assigns one presence slot to each domain value and an additional slot to \texttt{NULL}, when required. It forms a masked membership indicator, reduces each slot on the MXU, and emits the values whose presence bits are set. The output size is determined by the encoded domain rather than the input cardinality.

\para{Sort-based distinct}
The sort-based implementations order value tuples, identify boundaries between unequal adjacent tuples, and retain the first active row of each run. \opCompactSortDistinct{} first compacts the input when only a small fraction of its compiled rows are active, reducing the number of rows entering the sort. \opVMEMBitonicDistinct{} uses \texttt{jax.lax.sort} when the active tuple payload fits the calibrated VMEM budget. \opDirectSortDistinct{} applies \texttt{jnp.unique(size=...)} to the full compiled input as the general fallback. All three produce fixed-size output tensors with a validity mask identifying the unique rows.

\para{Grouped distinct}
For \texttt{COUNT(DISTINCT x)} grouped by keys $\mathbf{g}$, NTA treats $(\mathbf{g},x)$ as the distinct identity. It removes duplicate valid pairs and then counts the surviving values for each group. The intermediate may therefore contain nearly as many rows as the input, even when the final number of groups is small.

\subsection{Sort and Top-K} \label{sec:operators:topk}

Sort implementations differ in how much ordering they materialize and where they complete it. For \texttt{ORDER BY ... LIMIT K}, \opInKernelTopK{} masks invalid keys, extracts key--row-identifier pairs with \texttt{jax.lax.top\_k}, and gathers non-key payload columns only after the winners are known. \opGlobalSort{} uses \texttt{jnp.argsort} when the entire relation must be ordered. For a small statically bounded candidate set, \opHostHybridSort{} transfers only keys and row identifiers to the host for final ordering and fetches the selected payload columns lazily, limiting host--device traffic.

\section{TPU HW-Conscicous Compilation \& Optimization} \label{sec:compiler}
To address the shape and address dynamism described in Section~\ref{sec:dynamism}, NTA compiles a logical query plan into a statically shaped JAX/Pallas program through three steps. (1) Static shape planning uses catalog statistics to assign each intermediate a compiled cardinality bound, which determines its tile-aligned tensor shape and buffer allocation. (2) Physical operator selection uses these planned bounds, the physical properties available at each operator, and calibrated cost models to choose a TPU-native implementation for every logical operator. (3) Plan finalization groups the selected implementations into fusion-compatible producer--consumer regions and lowers the resulting physical plan to JAX/Pallas for XLA JIT compilation.

TPUs shift scheduling and memory management from hardware to software. When intermediate cardinalities, data movement, and operator schedules can be bounded and planned before execution, NTA exposes compact, statically bounded, and fusable work to XLA, enabling dense pipelines with high hardware utilization.


\subsection{Static Shape Planning} \label{sec:compiler:shape}

Static shape planning determines the tensor shape of every intermediate before XLA compilation. For each logical operator $o$, NTA distinguishes two compile-time quantities: an estimated output cardinality $\widehat{N}_o$, derived from catalog statistics and operator semantics, and a compiled cardinality bound $B_o \ge \widehat{N}_o$. The estimate supports cardinality estimation and cost modeling, whereas the compiled bound determines the intermediate's tile-aligned tensor shape and buffer allocation. The compiler also uses $B_o$ to check whether the resulting buffers fit in device memory and to account for padded work during physical operator selection.

\para{Base selections}
For a selection over a base relation with $N$ rows, histograms, most-common-value frequencies, and the number of distinct values (NDVs) provide an upper selectivity estimate $s_i^+$ for each predicate $i$. Joint statistics directly provide the combined estimate $s^+$ when available. Otherwise, NTA combines predicate estimates conservatively:
\begin{equation}
  s^+_{\land}=\min_i s_i^+,
  \qquad
  s^+_{\lor}=\min\left(1,\sum_i s_i^+\right).
\end{equation}
This avoids treating attribute independence as an upper bound. NTA derives the compiled bound for the selection as
\begin{equation}
  B_{\mathrm{sel}}=\operatorname{TileUp}\!\left(
  \min\!\left(N,\left\lceil \gamma N s^+\right\rceil\right)\right),
\end{equation}
where $\gamma > 1$ is a calibrated safety factor and \textsc{TileUp} rounds to the static tile size required by the selected implementation. When the required statistics are unavailable, NTA conservatively sets $s^+=1$.

\para{Bound propagation}
Starting from the base inputs, the compiler propagates compiled cardinality bounds bottom-up through the logical plan. Let $B_{\mathrm{in}}$ and $B_{\mathrm{out}}$ denote the compiled input and output bounds of an operator. A PK--FK join with referential integrity produces no more rows than its foreign-key input, giving the bound $B_{\mathrm{FK}}$. If the compiled PK input has bound $B_{\mathrm{PK}}$ and the layout guarantees a maximum fanout $k$, NTA uses the tighter bound
\begin{equation}
    B_{\mathrm{out}}=\min(B_{\mathrm{FK}},kB_{\mathrm{PK}}).
\end{equation}
A unique-build equijoin executed by \opSortedHashLookupJoin{} produces at most one match per probe row, so its output bound equals its probe-input bound.

For a grouped aggregation over keys $G_i$, NTA uses
\begin{equation}
    B_{\mathrm{out}}=\min\left(
    B_{\mathrm{in}},
    \prod_i\operatorname{NDV}(G_i)
    \right).
\end{equation}
Duplicate elimination uses the corresponding value-domain bound, while \texttt{ORDER BY ... LIMIT K} uses
\begin{equation}
    B_{\mathrm{out}}=\min(B_{\mathrm{in}},K).
\end{equation}

\para{Runtime validation}
Compiled cardinality bounds are guarded execution bounds rather than semantic worst-case bounds. Each bounded operator writes at most $B_o$ tuples to its statically shaped output and reports an overflow if its observed output cardinality would exceed $B_o$. The host accepts a query result only if all overflow checks succeed. If $N_o^{\mathrm{obs}} > B_o$, NTA discards the result, increases the violated bound, recomputes the downstream bounds, reruns physical operator selection, and recompiles the same logical plan. No such recompilation occurred in our evaluated workloads.

\subsection{Physical Operator Selection} \label{sec:compiler:decision}

\begin{figure*}[t]
  \centering
  \begin{tikzpicture}[
    x=1cm,
    y=1cm,
    font=\sffamily\scriptsize,
    >={Stealth[length=3.5pt,width=4.2pt]},
    noedge/.style={-{Stealth[length=3.5pt,width=4.2pt]}, draw=ntaFigureGray!60!ntaFigureInk,
      line width=.55pt, shorten >=1pt},
    yesedge/.style={-{Stealth[length=3.5pt,width=4.2pt]}, draw=ntaFigureBlue,
      line width=.70pt, shorten >=1pt},
    panel/.style={draw=ntaFigureHairline, fill=white, rounded corners=3.5pt, line width=.50pt},
    header/.style={draw=none, fill=ntaFigureBlueLight!48, rounded corners=2.5pt,
      minimum width=7.92cm, minimum height=.44cm},
    decision/.style={draw=ntaFigureGray, fill=white, text=ntaFigureInk, rounded corners=2pt,
      line width=.52pt, text width=2.52cm, minimum height=.48cm, inner sep=1.8pt,
      align=center, font=\sffamily\scriptsize},
    leaf/.style={draw=ntaFigureBlue!70, fill=ntaFigureBlueLight!55, text=ntaFigureInk, rounded corners=2pt,
      line width=.58pt, text width=2.45cm, minimum height=.48cm, inner sep=1.8pt,
      align=center, font=\sffamily\scriptsize\bfseries},
    title/.style={font=\sffamily\footnotesize\bfseries, text=ntaFigureInk},
    legend/.style={font=\sffamily\tiny, text=ntaFigureGray}
  ]

  \node[panel, minimum width=8.35cm, minimum height=4.72cm, anchor=north west] (jp) at (0,0) {};
  \node[header] at (4.17,-.30) {};
  \node[title, anchor=west] at (.22,-.30) {(a) Join};
  \node[legend, anchor=east] at (8.12,-.30)
    {\textcolor{ntaFigureBlue}{$\rightarrow$ yes}\quad \textcolor{ntaFigureGray!60!ntaFigureInk}{$\downarrow$ no}};

  \node[decision, text width=2.40cm] (j1) at (1.55,-0.78) {1:1 join with identical row-aligned keys?};
  \node[decision, text width=2.40cm] (j2) at (1.55,-1.48) {Clustered PK--FK with bounded fanout $k$?};
  \node[decision, text width=2.40cm] (j3) at (1.55,-2.18) {Existing PK--FK lookup index?};
  \node[decision, text width=1.85cm] (j4) at (4.18,-2.18) {SparseCore feasible \& above crossover?};
  \node[decision, text width=2.40cm] (j5) at (1.55,-3.58) {Unique-build equijoin?};

  \node[leaf] (jka)  at (6.85,-0.78) {Key-Aligned SIMD Join};
  \node[leaf] (jfr)  at (6.85,-1.48) {Fanout-Reshape Join};
  \node[leaf] (jsg)  at (6.85,-2.18) {SparseCore Gather Join};
  \node[leaf] (jvpu) at (6.85,-2.88) {VPU Index Lookup Join};
  \node[leaf] (jsh)  at (6.85,-3.58) {Sorted-Hash Lookup Join};

  \draw[yesedge] (j1.east) -- (jka.west);
  \draw[noedge]  (j1.south) -- (j2.north);
  \draw[yesedge] (j2.east) -- (jfr.west);
  \draw[noedge]  (j2.south) -- (j3.north);
  \draw[yesedge] (j3.east) -- (j4.west);
  \draw[noedge]  (j3.south) -- (j5.north);
  \draw[yesedge] (j4.east) -- (jsg.west);
  \draw[noedge]  (j4.south) |- (jvpu.west);
  \draw[yesedge] (j5.east) -- (jsh.west);

  \node[panel, minimum width=8.35cm, minimum height=4.72cm, anchor=north west] (ap) at (8.70,0) {};
  \node[header] at (12.87,-.30) {};
  \node[title, anchor=west] at (8.92,-.30) {(b) Aggregation};
  \node[legend, anchor=east] at (16.82,-.30)
    {\textcolor{ntaFigureBlue}{$\rightarrow$ yes}\quad \textcolor{ntaFigureGray!60!ntaFigureInk}{$\downarrow$ no}};

  \node[decision, text width=2.58cm] (a1) at (10.30,-0.78) {Small dense group domain with MXU support?};
  \node[decision, text width=2.58cm] (a2) at (10.30,-1.48) {Histogram working set fits VMEM?};
  \node[decision, text width=2.58cm] (a3) at (10.30,-2.18) {Multi-key with small packed domain?};
  \node[decision, text width=2.58cm] (a4) at (10.30,-2.88) {Input ordered by group key?};
  \node[decision, text width=2.58cm] (a5) at (10.30,-3.58) {Clustered groups with bounded fanout $k$?};

  \node[leaf] (aoh)  at (15.35,-0.78) {One-Hot MatMul Aggregation};
  \node[leaf] (avh)  at (15.35,-1.48) {VMEM Block Histogram};
  \node[leaf] (arp)  at (15.35,-2.18) {Radix-Packed Aggregation};
  \node[leaf] (asr)  at (15.35,-2.88) {Sorted-Run Aggregation};
  \node[leaf] (afr)  at (15.35,-3.58) {Fanout-Reshape Aggregation};
  \node[leaf, text width=2.58cm] (aseg) at (10.30,-4.28) {Segment Reduction};

  \draw[yesedge] (a1.east) -- (aoh.west);
  \draw[noedge]  (a1.south) -- (a2.north);
  \draw[yesedge] (a2.east) -- (avh.west);
  \draw[noedge]  (a2.south) -- (a3.north);
  \draw[yesedge] (a3.east) -- (arp.west);
  \draw[noedge]  (a3.south) -- (a4.north);
  \draw[yesedge] (a4.east) -- (asr.west);
  \draw[noedge]  (a4.south) -- (a5.north);
  \draw[yesedge] (a5.east) -- (afr.west);
  \draw[noedge]  (a5.south) -- (aseg.north);

  \end{tikzpicture}
  \caption{Physical operator selection for joins and grouped aggregations. Blue rightward branches denote satisfied conditions; gray downward branches continue to the next alternative. Conditions and leaves correspond to Table~\ref{tab:rewrite_map}.}
  \label{fig:operator_selection}
\end{figure*}
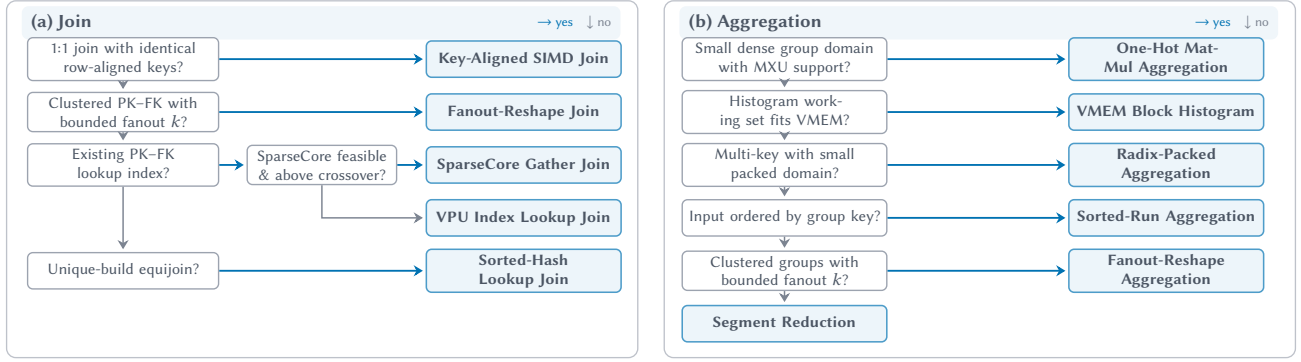

After static shape planning determines the compiled cardinality bounds, NTA selects one physical implementation for each logical operator. Selection first applies hard constraints: an implementation is considered only when its required constraints and input layout are available and its working set fits the target TPU. Performance-dependent branches then use crossover rules calibrated for the target hardware. Figure~\ref{fig:operator_selection} shows the resulting decision paths for joins and grouped aggregations.

\para{Join and aggregation choices}
Figure~\ref{fig:operator_selection} applies the required properties in Table~\ref{tab:rewrite_map}. For joins, NTA first exploits identical row alignment or clustered PK--FK fanout. An existing PK--FK lookup index enables either \opVPUIndexLookupJoin{} or \opSparseGatherJoin{}, with the hardware profile and calibrated probe-size crossover selecting the backend. Without an index, a unique-build equijoin enables \opSortedHashLookupJoin{}. For aggregation, NTA considers a small dense domain, VMEM-resident histogram state, a small packed multi-key domain, existing key order, and clustered bounded fanout in turn. Large or sparse group domains use \opSparseSegmentReduction{}.

\para{Other operators}
Distinct and Sort/Top-K use the same property-first procedure. For \texttt{DISTINCT}, a small bounded integer domain selects \opMXUOneHotDistinct{}; otherwise, a low active-row fraction selects \opCompactSortDistinct{}, a working set that fits VMEM selects \opVMEMBitonicDistinct{}, and \opDirectSortDistinct{} is the general fallback. For \texttt{ORDER BY}, a bounded \texttt{LIMIT} or \texttt{OFFSET} selects \opInKernelTopK{}. Without such a bound, a small host-finalization boundary enables \opHostHybridSort{}; remaining cases use \opGlobalSort{}. Selections and projections have no independent physical implementation and fuse into the selected consumer unless materialization is required.

\para{Selection thresholds}
Terms such as ``small'' and ``above crossover'' refer to target-specific thresholds rather than universal constants. Some choices follow directly from input properties: NTA selects SparseCore
Gather for PK--FK joins with a prebuilt index and Sorted-Hash Lookup when no such index is available, as shown in Figure~\ref{fig:join_benchmark}. When both VPU and SparseCore indexed joins are applicable, the v5p policy switches to SparseCore at 50K
probe rows. In the 600M-row aggregation sweep, NTA selects One-Hot MatMul through $K=256$; at $K=1024$, the $N\times K$ representation
exceeds device memory, and NTA switches to Segment Reduction through the largest evaluated domain, $K=2^{20}$. For Distinct, One-Hot admits at most 128 value slots, Compact-Sort requires an active-row fraction at most 20\% over an input of at least 100K rows, and VMEM sort admits at most
8\,MiB of active payload. In-Kernel Top-K requires $K\le4096$, while Host-Hybrid Sort admits at most 20K candidates.

\subsection{Plan Finalization and Lowering}\label{sec:compiler:lowering}
After selecting a physical implementation for each operator, NTA assembles the selected operators into a complete physical plan. It fuses non-materializing operations, such as selections and projections, into the adjacent operators that consume them and constructs any composite keys required by joins and grouped aggregations. This removes unnecessary intermediate tensors and keeps adjacent computation together where possible. NTA then applies the tile-aligned tensor shapes determined in Section~\ref{sec:compiler:shape}, propagates validity masks through the plan, and adds overflow checks to operators with data-dependent output cardinalities. Finally, it lowers regular tensor computation to JAX and custom data-movement kernels to Pallas. XLA compiles the resulting statically shaped program, allocates its buffers, and schedules its execution.

\section{Orchestration at Extreme Scale}\label{sec:orchestration}
While single-device execution provides a foundation, processing hundreds of gigabytes to terabytes (e.g., SF1000) requires coordinating query execution across multiple TPU chips. This section describes how NTA distributes physical query plans across TPU devices and schedules concurrent queries under the TPU runtime's execution constraints.

\para{Scaling Out with JAX Distributed}
NTA uses JAX's distributed runtime (\texttt{jax.distributed}) and XLA's
GSPMD partitioner~\cite{xu2021gspmd} to execute a physical query plan across a TPU slice. The devices in the assigned slice form a logical JAX device mesh used for both SPMD compilation and data (relation) placement. XLA compiles the physical query plan for this execution layout, and each device processes the relation partitions assigned to it.

A naive execution strategy partitions each relation independently and relies on XLA to route indexed lookups to the device containing the referenced tuple. This can turn PK--FK joins into distributed joins that require communication for every lookup. Retaining global tuple identifiers also requires 64-bit indices once a relation exceeds the signed 32-bit row range, doubling index storage and index-read traffic on TPUs.
NTA instead preserves partitioning metadata and row alignment across physical operators. Large relations are partitioned across devices, whereas small relations are replicated when they fit in device memory. Relations connected by high-volume PK--FK joins are co-partitioned when their existing layouts are compatible. The corresponding indexes use partition-local signed 32-bit tuple identifiers, allowing each device to evaluate the indexed join using only its local partitions.

Partition-local identifiers can also serve as group identifiers when the grouping attributes are aligned with the partitioning scheme. In this case, each device computes a local aggregation without exchanging input tuples. Each partition carries a stable partition identifier, and a tuple identifier that appears in the query result is represented as
$(\mathit{partition\_id}, \mathit{local\_id})$ and decoded during result materialization. Lookups into replicated relations are also local. Joins whose
inputs are neither replicated nor compatibly partitioned require a repartitioning exchange.

Each device executes the compiled query over its local relation partitions. Scans, selections, co-partitioned indexed joins, and local aggregations proceed without communication. Data exchange is required only for repartitioning and
for combining partial results into the final result relation.

\para{Concurrency and Multi-Tenant Scheduling}
While the TPU excels at raw throughput, managing concurrent queries in a unified compilation model exposes static execution constraints. In our current JAX/TPU deployment, the host submits query executables to a serialized, First-In-First-Out (FIFO) instruction queue. Consequently, unlike CPU databases that time-slice concurrent analytical queries, NTA's average query latency scales linearly with the number of concurrent tenants, even though aggregate cluster throughput remains constant.

\section{Evaluation} \label{sec:experiments}
We evaluate NTA to validate three core claims: (1) our hardware-conscious algorithms effectively reduce TPU memory-bound stalls, (2) our hardware-conscious local physical planner selects performant JAX and Pallas implementations, and (3) our JAX distributed orchestration scales effectively to extreme bounds.

\subsection{Experimental Setup}
\label{sec:eval:setup}

\para{Benchmarks} We evaluate NTA on TPC-H~\cite{TPCHSpec} and ClickBench~\cite{clickbench2024}. We run all 22 TPC-H queries at SF1, SF100, and SF1000 for the main benchmark experiments. We also run 
ClickBench queries (standard 100M row table) for generalization \& limitations study (discussed in Section~\ref{sec:generalization}).\par

\para{Baselines \& Setup} We compare DuckDB, Direct JAX, and NTA. \textbf{DuckDB} is our SIMD-vectorized, in-memory CPU baseline, a widely-accepted analytics query processing baseline~\cite{CoddSpeed,yogatama2024lancelot,wu2025terabyte,yuan2024vortex}. \textbf{Direct JAX} follows TQP~\cite{TQP}'s tensor query mapping using JAX/XLA on TPU, without our native TPU kernels and intelligent kernel selection;  \textbf{NTA} is the complete system. Direct JAX isolates NTA's contribution on the same TPU hardware. 
All runs use isolated Linux containers. All runs use isolated Linux containers. DuckDB runs on a dual-socket Google Cloud host equipped with two Intel Xeon Platinum 8481C processors (112 physical cores, 224 vCPUs via 2-way SMT) and up to 400 GB of memory. DuckDB is configured with 224 worker threads, matching the 224 hardware execution contexts (its default auto-detected concurrency) without oversubscribing the CPU. TPU runs use v5p unless noted, with selected single-chip runs repeated on v7p; Direct JAX and NTA use identical allocations. Each host supports at most four chips, so $D$ chips span $\lceil D/4 \rceil$ hosts.\par

\para{Measurements} All measurements are warm: input data is resident in memory and TPU executables are precompiled. We time query execution only, excluding data generation, loading, host-to-device transfer, and JIT compilation. We report the median latency of five runs and speedup over DuckDB. We summarize each query set by geometric-mean speedup and total-runtime speedup, the ratio of summed DuckDB to system runtimes.\par

\subsection{Macrobenchmarks (SF1 \& SF100)}
\begin{figure*}[t]
  \centering
  \captionsetup{skip=1pt}
  \includegraphics[width=\textwidth]{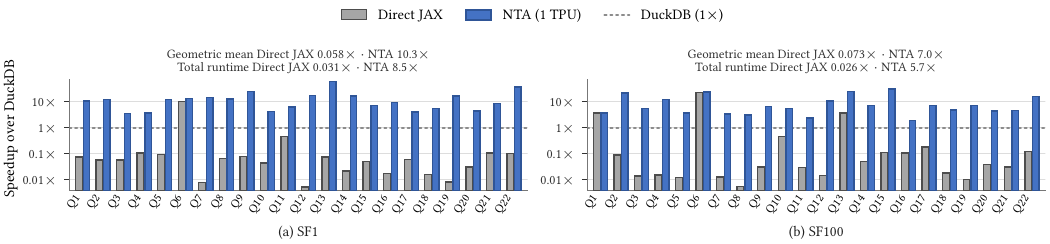}
  \caption{Per-query speedup over DuckDB on 1 TPU v5p chip at (a) SF1 and (b) SF100.}
  \Description{Two side-by-side log-scale panels compare Direct JAX and NTA per-query speedup over DuckDB on one TPU v5p chip at SF1 and SF100. Each panel reports both geometric-mean and total-runtime speedup.}
  \label{fig:tpch_sf1_speedup}
\end{figure*}

Figure~\ref{fig:tpch_sf1_speedup}(a) reports per-query speedup over DuckDB at SF1. Direct JAX achieves a geometric-mean speedup of $0.058\times$ and outperforms DuckDB only on Q6, whereas NTA accelerates every query by $3.5$--$61.0\times$, achieving $10.3\times$ geometric-mean and $8.5\times$ total-runtime speedup. At SF100 (Figure~\ref{fig:tpch_sf1_speedup}(b)), Direct JAX achieves $0.073\times$ geometric-mean and $0.026\times$ total-runtime speedup, outperforming DuckDB only on Q1, Q6, and Q13. NTA accelerates every query by $1.9$--$30.5\times$ and achieves $7.0\times$ geometric-mean and $5.7\times$ total-runtime speedup.

\para{Sources of improvement} Q6 is the only Direct JAX query faster than DuckDB at both scales because its selective scan and scalar reduction already map to a compact fused tensor program. At SF100, Q1 and Q13 also cross the baseline, but most Direct JAX join and grouped-reduction queries remain far slower. NTA retains the compact Q6 path and additionally regularizes the joins and grouped reductions that dominate the remaining queries. Selective dimension predicates become masks before fact-table access, indexed payloads use masked SparseCore gathers, and bounded or small-domain groups use reshape, histogram, or dense reduction rather than global scatter. These choices are especially effective for Q9 ($25.3\times$ at SF1), whose multi-hop joins reuse compact lookup payloads, and for Q12 ($16.9\times$), whose selective order lookup feeds a two-group reduction. Q13 and Q22 obtain the largest SF1 gains ($61.0\times$ and $36.7\times$): Q13 scans dictionary tokens and reduces directly into a 61-bin customer-count distribution, while Q22 extracts phone-code bytes and aggregates the surviving customers into a small dense domain.

\para{Residual bottlenecks} The benefit depends on how early the selected implementation collapses the working set. At SF100, Q15 ($30.5\times$), Q13 ($25.4\times$), and Q6 ($23.5\times$) retain compact aggregate state while their input scans grow. In contrast, Q1, Q5, Q7, and Q8 still stream much of \textit{lineitem} and several payload columns, leaving them bounded by HBM bandwidth at $3.1$--$3.7\times$. Q16 has the smallest gain ($1.9\times$) because substring predicates, distinct supplier counting, and a full ordering over tens of thousands of output groups leave substantial per-row and high-cardinality work after filtering; Q11 similarly scans \textit{partsupp}, forms per-part state, and sorts its surviving groups. The takeaway is that TPU-native operators remove irregular-access overhead, but they cannot eliminate bytes that the query must scan or large result state that it must materialize.

\begin{figure*}[t]
  \centering
  \captionsetup{skip=1pt}
  \includegraphics[width=\textwidth]{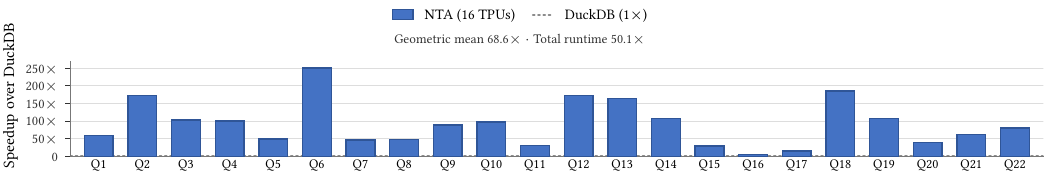}
  \caption{Per-query speedup over DuckDB on 16 TPU v5p chips at SF1000.}
  \Description{A linear-scale bar chart starting at zero shows per-query NTA speedup over DuckDB for all 22 TPC-H queries at SF1000. NTA uses 16 TPU v5p chips and accelerates every query. The figure reports 68.6-times geometric-mean and 50.1-times total-runtime speedup.}
  \label{fig:tpch_sf1000_speedup}
\end{figure*}

\subsection{Extreme Scale (SF1000)}
Figure~\ref{fig:tpch_sf1000_speedup} reports SF1000 performance using 16 TPU v5p chips against DuckDB. NTA accelerates every query by $5.5$--$250.7\times$, achieving $68.6\times$ geometric-mean speedup. The complete query set finishes in 1.613 seconds, down from 80.86 seconds, yielding $50.1\times$ total-runtime speedup.

\para{Scale-out behavior} Row-wise co-sharding keeps scans, masks, indexed gathers, and local reductions partition-local; devices exchange only bounded aggregate state or Top-K candidates. Q6 therefore reaches $250.7\times$: each chip performs a contiguous scan and contributes one scalar partial sum. Q2, Q12, and Q13 achieve $164.8$--$172.0\times$ because selective predicates sharply reduce the fact-side work and the final Top-K or group state is small. Q3 and Q18, which are less dominant on one chip, reach $103.2\times$ and $185.3\times$ because clustered per-order reductions execute independently within each shard and require only a small final Top-K merge.

\para{Scale-out limits} Queries with large per-shard state or work that does not shrink with row partitioning remain the tail. Q16 improves by only $5.5\times$ because string matching, high-cardinality distinct counting, and final ordering are still performed within every shard; Q17 reaches $15.4\times$ because its correlated per-part average requires variable-fanout child reads and multiple reductions. Q11, Q15, and Q20 also retain per-key state or dependent filtering passes, limiting them to $29.8$--$40.0\times$. Consequently, the $50.1\times$ total-runtime gain is lower than the $68.6\times$ geometric mean: the remaining long-running, state-heavy queries receive less benefit and dominate elapsed time. Even so, the slowest query remains $5.5\times$ faster, showing that NTA's static programs scale without shifting the bottleneck to inter-device exchange.

\subsection{Ablation Study}
\begin{figure}[!ht]
  \centering
  \includegraphics[width=\linewidth]{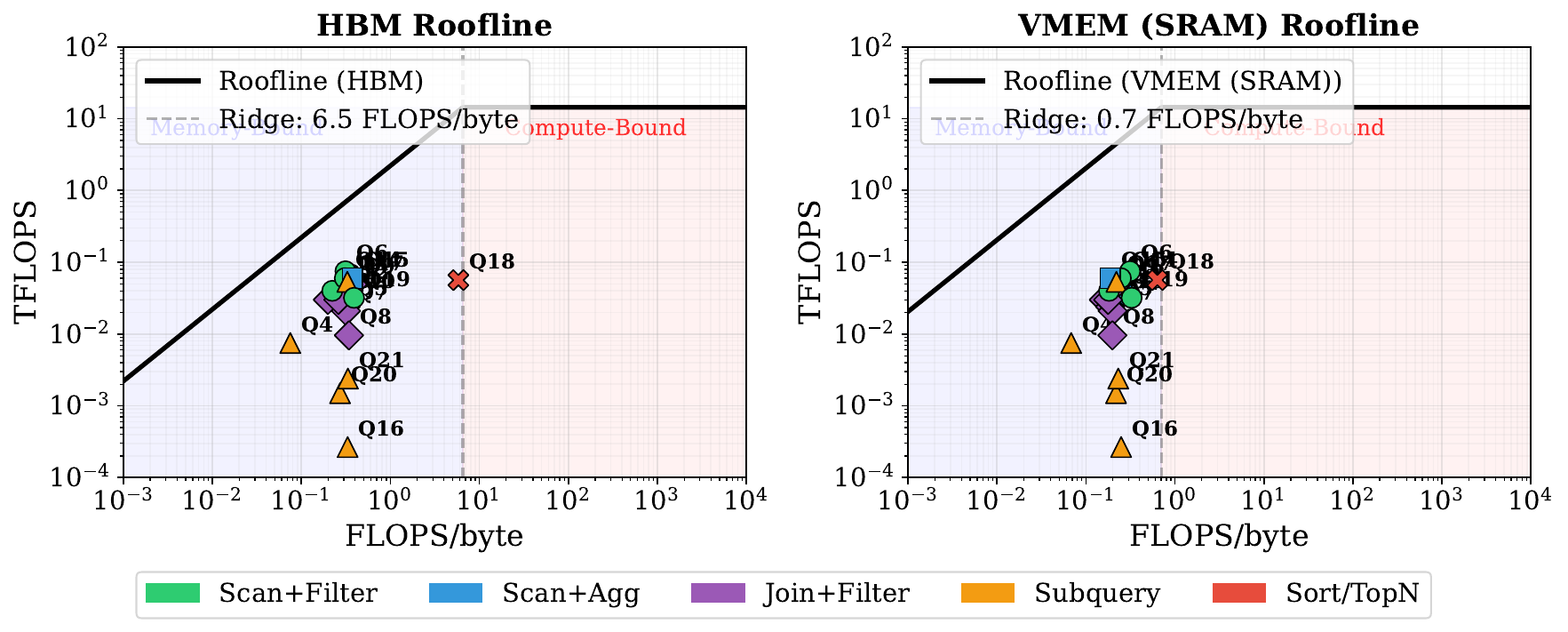}
  \caption{Dual roofline analysis for the optimized NTA on TPC-H (SF1).}
  \label{fig:ntdb_roofline}
\end{figure}

\para{Roofline Analysis}
To ground our operator-level performance, we examine our TPC-H query execution against the TPU's Roofline model in Figure~\ref{fig:ntdb_roofline}. 
The left panel shows the HBM Roofline (ridge point at 6.5 FLOPS/byte). Because relational algebra streams arrays, almost all queries (Scan, Aggregation, Join) heavily populate the memory-bound region (0.1--1.0 FLOPS/byte) when constrained by HBM. Only heavy sorting operators (e.g., Q18) approach the compute-bound region. 
Relative to VMEM, many of our queries transition decisively to the ridge point or fully into the compute-bound region. This demonstrates our TPU-native execution perfectly transitions inherently memory-bound analytical operations into tightly fused tensor math limited only by raw compute.

\begin{figure}[!ht]
    \centering
    \includegraphics[width=\linewidth]{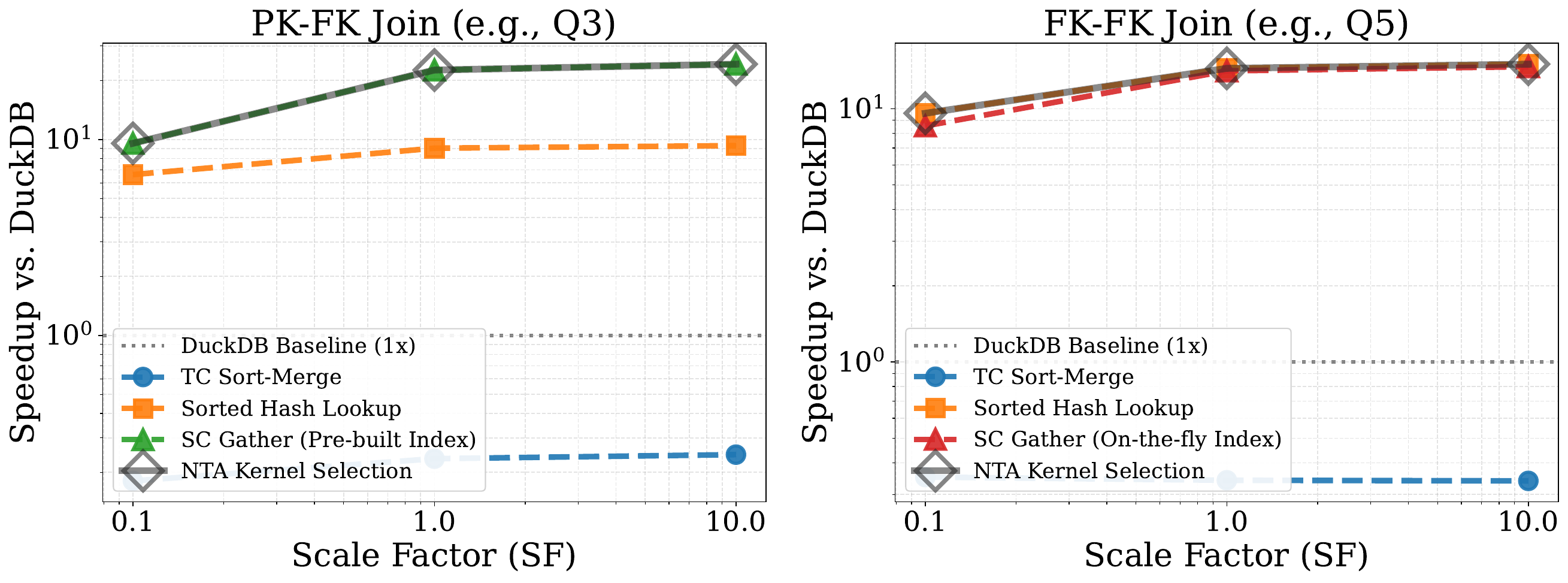}
    \caption{Single-chip join (PK-FK, FK-FK) benchmark across scale factors.}
    \label{fig:join_benchmark}
\end{figure}

\para{Intelligent Join Kernel Selection} Figure~\ref{fig:join_benchmark} evaluates single-chip join kernel performance across scale factors, SF0.1 to SF10 for two join patterns. For primary-foreign key (PK-FK) joins (e.g., Q3) foreign key indices can be pre-built and mapped directly. NTA detects and optimally routes the execution to the \texttt{Pallas SC Gather} kernel, which leverages the SparseCore to achieve fast (< ms) lookups and up to $41.4\times$ speedup over DuckDB at SF10. 
For FK-FK joins (e.g., Q5) where pre-built index maps are unavailable, NTA dynamically switches to the  \texttt{Sorted Hash Lookup Join}. This intelligent selection avoids the heavy overhead of sorting algorithms, outperforming the na\"{\i}ve \texttt{TC Sort-Merge Join} baseline by over $15\times$ while maintaining a robust $18.4\times$ speedup over the CPU baseline at SF10.

\begin{figure}[!ht]
    \centering
    \includegraphics[width=0.8\linewidth]{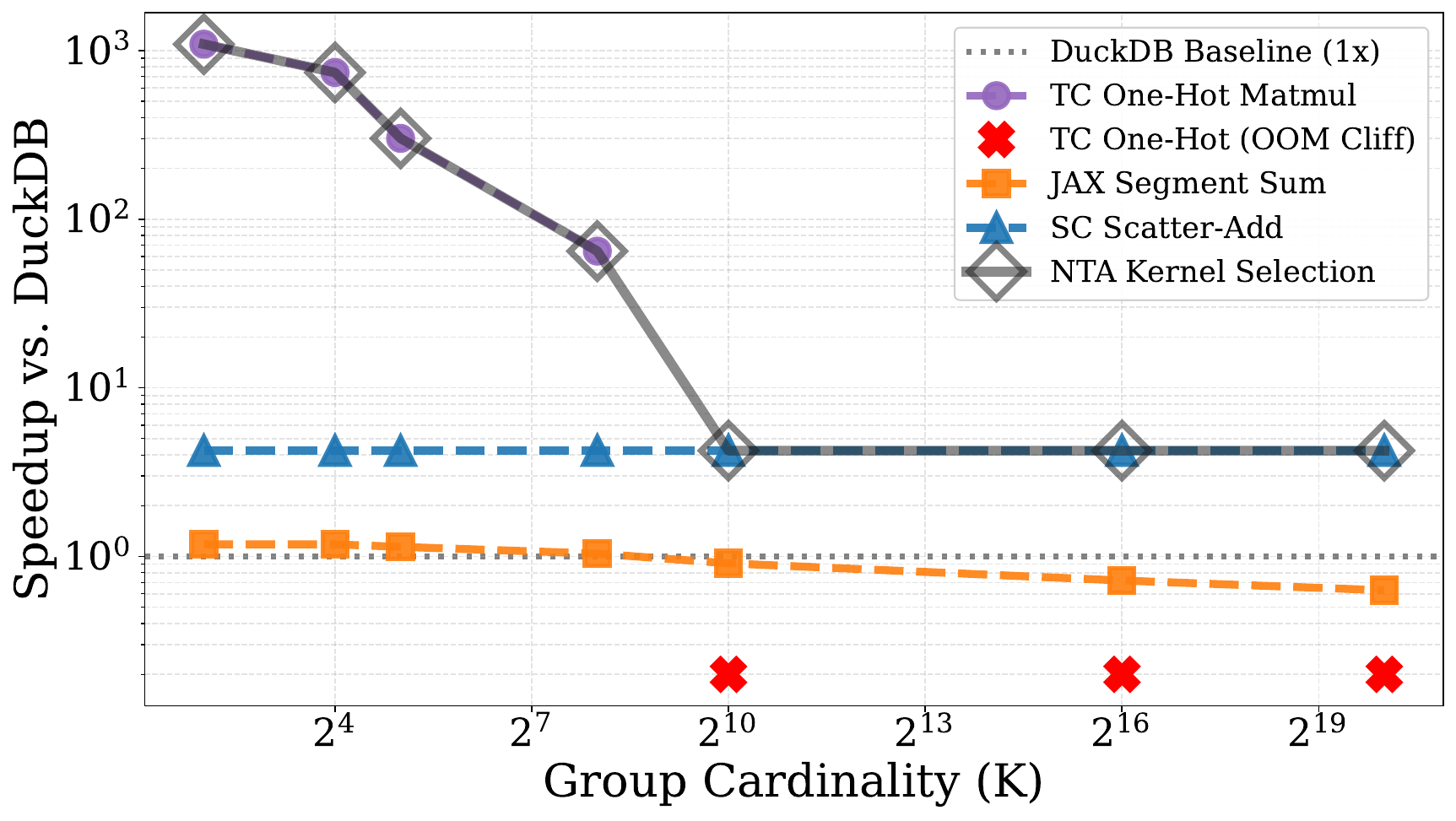}
    \caption{Single-chip aggregation benchmark varying group cardinalities ($K$).}
    \label{fig:agg_benchmark}
\end{figure}

\para{Intelligent Aggregation Kernel Selection} Figure~\ref{fig:agg_benchmark} illustrates how aggregation kernels behave under varying group cardinalities, $K$, for a fixed 600M-row workload. For low-cardinality ($K \le 256$), the \texttt{TC One-Hot Matmul} kernel utilizes the systolic MXU array, achieving over $1000\times$ speedup at $K=4$. However, because it materializes an $N\times K$ dense matrix, it encounters HBM Out-of-Memory (OOM) errors for $K \ge 1024$. At higher cardinalities, NTA instead selects \opSparseSegmentReduction{}, which uses XLA scatter-add through \texttt{jax.ops.segment\_sum} and avoids materializing the $N\times K$ matrix. This hardware-aware selection prevents OOM failures, achieving a robust $4.2\times$ speedup over the CPU baseline.

\begin{figure}[!ht]
    \centering
    \includegraphics[width=0.8\linewidth]{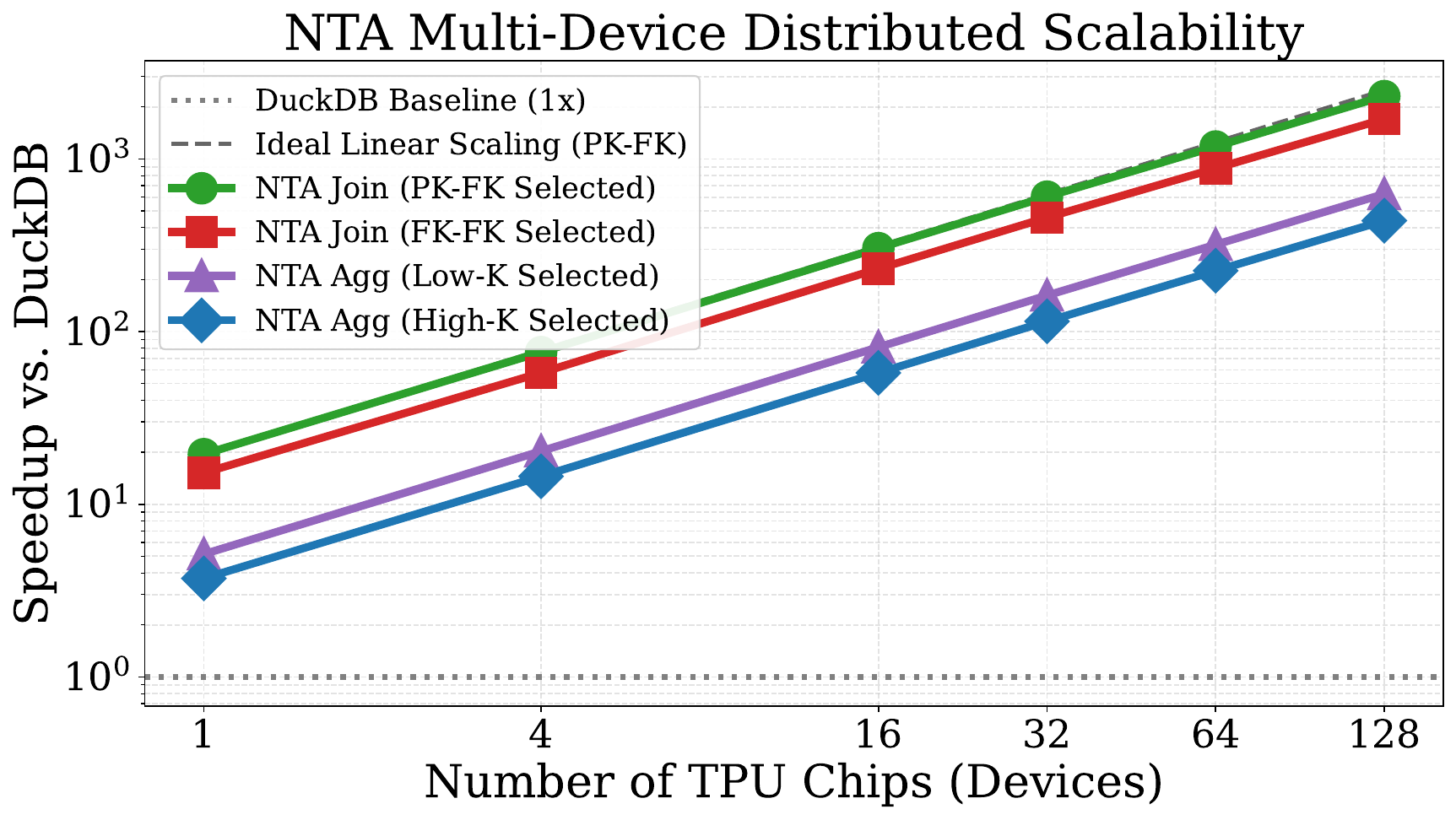}
    \caption{NTA scalability using JAX distributed (SPMD) up to 128 TPU v5p chips.} 
    \label{fig:tpu_device_scalability}
\end{figure}

\para{Multi-Device Scaling} Figure~\ref{fig:tpu_device_scalability} illustrates the distributed scalability of NTA across the TPU pod slices (1 to 128 TPU chips with fast ICI links). JAX/XLA SPMD shards the data row-wise across the available devices, minimizing interconnect bottlenecks.
Both the selected join kernels (PK-FK SC Gather, FK-FK Sorted Hash), the selected aggregation kernels (Low-$K$ One-Hot-Matmul, High-$K$ Segment Reduction) exhibit near-linear distributed speedup up to 128 devices.
This demonstrates the effectiveness of our hardware conscious kernel design and selection.

\begin{figure}[!th]
    \centering
    \includegraphics[width=0.8\linewidth,height=0.5\linewidth]{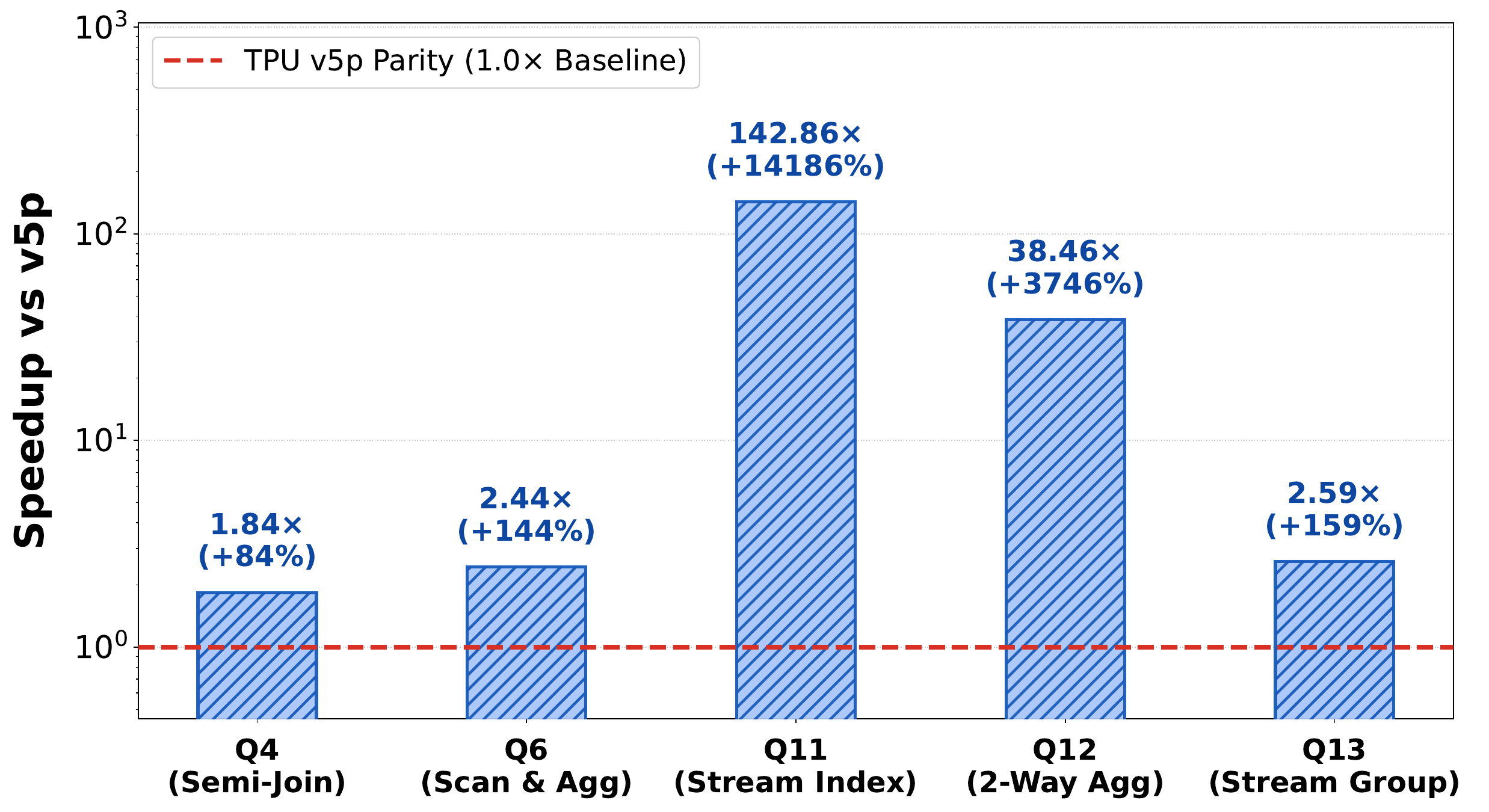}
    \caption{TPU v7 speedups over TPU v5p (streaming \& scan-heavy queries, TPC-H SF100)}
    \label{fig:tpch_v7_gain}
\end{figure}

\para{Scan \& Streaming Microbenchmark} We evaluate streaming and scan-heavy queries (Q4, Q6, Q11, Q12, Q13 in TPC-H SF100). They expose performance ceilings due to HBM bandwidth, DMA engines, and scratchpad (SMEM, VMEM) capacities. The TPU v5p vs. v7 ablation illustrates that the latest v7~\footnote{TPU v7 offers twice the HBM capacity and more than $2.6\times$ the HBM bandwidth, see \url{https://docs.cloud.google.com/tpu/docs/tpu7x}} with much larger bandwidth and capacity can unlock much stronger performance through both hardware improvements (e.g., 2.7x HBM bandwidth increase) and software kernel redesigns.
For instance, Q11 and Q12 can stage entire dimension filters within v7's expanded registers/SMEM, avoiding extra HBM roundtrips. Similarly, the left-outer join results (order count per customer) in Q13 can be pipelined and resident on-chip, eliminating extra HBM roundtrips and multi-kernel launches for the separate aggregations.

\section{Discussion \& Future Work} \label{sec:discussion}
\subsection{Generalization \& Limitations} \label{sec:generalization}
\begin{figure}[!th]
  \centering
  \captionsetup{skip=1pt}
  \includegraphics[width=\columnwidth]{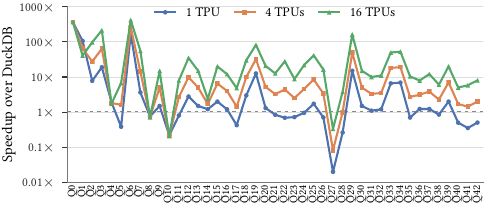}
  \caption{Warm-execution speedup over DuckDB for each ClickBench query using 1, 4, and 16 TPUs.}
  \Description{A log-scale plot shows per-query ClickBench speedup over DuckDB using 1, 4, and 16 TPUs.}
  \label{fig:workload_generalization}
\end{figure}

We evaluate NTA on ClickBench, a 100M-row flat-table analytics benchmark (Figure~\ref{fig:workload_generalization}). 
To isolate physical TPU execution from high-level optimization, NTA ingests fixed logical plans directly from DuckDB, delegating orthogonal logical rewrites (e.g., join reordering, CSE, factorized aggregation pushdown) upstream to focus on physical operator selection and tile scheduling.

Across ClickBench, performance is governed less by query complexity than by how early execution collapses into compact, statically bounded, and fusable state. 
Queries reducing to metadata lookups, scalar reductions, or small histograms scale exceptionally well: Q0 leverages row-count metadata, Q3 and Q6 perform scalar reductions, and Q29 computes 90 derived sums from a single base sum and row count, achieving up to $417\times$ speedup. 
Overall, a single TPU delivers a $1.79\times$ geometric-mean speedup, outperforming DuckDB on 26 of 43 queries ($4.36\times$ geomean on these wins); 4 TPUs achieve $5.17\times$ (winning 39 queries), and 16 TPUs reach $14.60\times$ (winning 40 queries). 
This demonstrates that NTA's benefits generalize beyond relational joins (TPC-H) to large-scale flat-table aggregations with bounded output domains.

In contrast, performance bottlenecks occur when intermediate states cannot fit on-chip: Q8 and Q10 compute exact distinct-user counts across grouping keys, requiring table-scale resolution of composite identities prior to reduction. 
Similarly, Q27 scans 100M rows with an unbounded URL key domain that exceeds on-chip scratchpad (VMEM) capacity, forcing histogram updates into latency-heavy HBM round-trips.

Variable string lengths and highly selective string matching (e.g., texttt{LIKE}, \texttt{regex}). This is challenging because the underlying static tensor shapes would require shorter strings to be right-padded to the worst-case bounds.
This padding and worst-case bound static tensor representation incur severe memory and bandwidth waste. When average string lengths are much shorter than the maximum, the majority of materialized tensors consist of zero-padding bytes that saturate scarce HBM bandwidth and on-chip scratchpad memory (VMEM). 
Furthermore, TensorCore VPU vector lanes waste significant compute throughput comparing these padded positions. We see this variable string lengths challenge as a future co-design opportunity for us.

Overall, NTA explicitly binds to JAX/XLA and Pallas kernels to bring about maximum performance from TPU and its various computing units. JAX/XLA also provides backend portability to different HW backends.
Although writing performant Pallas kernels traditionally requires manual tuning, emerging agentic database coding systems~\cite{wehrstein2026bespoke, lao2026gendb} offer a compelling solution path. By leveraging specialized database coding agents, systems can automatically synthesize, optimize, and specialize JAX/XLA and Pallas query kernels directly for TPU microarchitectures. 

\subsection{HW \& SW Co-Design Opportunities}
Re-engineering database operations for TPUs highlights residual performance gaps stemming from data-dependent dynamism, from \textit{redundant work} from shape dynamism (padded rows and masked SIMD lanes) and \textit{memory stalls} from address dynamism (fusion breaks at HBM roundtrips). To bridge these gaps without sacrificing the TPU's highly optimized static (compiler-based) execution model, we propose four critical hardware co-design directions:


\para{Dynamic Tile-Skipping} Static shape bounding forces VPUs to schedule and evaluate fully invalid or padded tiles. Lightweight hardware mask-evaluation units that inspect tile validity metadata prior to dispatch could dynamically bypass dead tiles, conserving memory bandwidth and freeing execution cycles for active data.

\para{Compaction of Sparse Vectors} Another source of sparsity is originated in the processing of the data, at the tuple level. Prior operators eliminate some of the tuples, which may not be consecutive in memory, leading to a finer-granular sparsity. For this case, compacting the data, locally near the VPU units, to have dense vectors would be a very useful hardware feature.

\para{Smart Caching} Unlike the hardware-managed caches seen in CPUs and GPUs, TPUs provide kernel-managed scratchpad memories (SPMEM/VMEM).
Prior work~\cite{cao2023gpu} has noted that cache evictions of hot data leads to significant performance drops.
For hardware-managed caches, this is unavoidable, as accessing transient data (e.g., tuples) pollutes the cache, evicting hot data (e.g., hash table entries) and requiring extra HBM accesses.
With TPUs, managing the scratchpad becomes a critical optimization point, allowing the database designer to ensure that frequently accessed data is accessed at a high bandwidth and low latency, reducing stalls.

\noindent\textbf{Hardware-Level Virtual Partitioning:} Currently, there is a lack of hardware partitioning and isolation on TPUs, enforcing strict sequential execution of query kernels and preventing efficient multi-tenancy. Future TPUs could benefit from virtual partitioning and hardware supported isolation~\cite{cao2023gpu}, to improve aggregate throughput and hardware utilization under concurrent workloads (Section~\ref{sec:orchestration}).
\vspace{-0.2cm}
\subsection{End-to-End Semantic Queries}
Emerging semantic AI query engines~\cite{liu2025palimpzest,chung2026architecting,patel2026towards,chung2026100x,ha2026cadenza} extend relational queries with \textit{semantic operators} (e.g., filters and aggregations evaluated implicitly via LLM inference). Today, these pipelines suffer from severe data transfer penalties as they constantly move across the host-to-accelerator boundary, alternating between CPU-based relational logic and LLM model inference on accelerator hardware. By deploying NTA, the entire semantic AI query (relational operators and LLM invocations) can be compiled into a unified JAX program that runs natively on the TPU end-to-end. We view this unification and elimination of the execution boundary as a promising application of our work, enabling the next generation of AI-native database engines~\cite{chung2026ai}.

\section{Related Work}

\para{Hardware-Accelerated Query Processing}
Relational query acceleration on massively parallel hardware has historically centered on GPUs. 
Early engines leveraged GPU SIMT architectures and hardware-managed caches to mask memory latency~\cite{GPUDB, GPUCPUStudy}, while subsequent systems integrated JIT compilation into heterogeneous CPU-GPU runtimes~\cite{Paul2021, Chrysogelos2019, Doraiswamy2023}. 
Concurrently, Tensor Query Processing (TQP)~\cite{TQP} proved that relational algebra can execute atop tensor computation runtimes—a paradigm now commercially deployed in Microsoft Fabric~\cite{CoddSpeed} with substantial speedups over CPU engines. Recently, modular GPU-native accelerators have emerged: Sirius~\cite{Sirius2026} accelerates existing engines (e.g., DuckDB, Apache Doris) via drop-in CUDA-X libraries (\texttt{libcudf}, RMM) for up to $10\times$ speedups, while Theseus~\cite{Theseus2025} optimizes distributed GPU query execution across high-speed interconnects.

However, these frameworks remain GPU-centric. 
While initial studies evaluated TPUs~\cite{TCUDB,TQP}, they largely treated the TPU as a black-box GPU drop-in, ignoring the severe HBM stalls and fusion breaks caused by the hardware's statically scheduled execution model. 
In contrast, NTA re-engineers relational query execution specifically for TPU hardware, directly co-designing physical operators around SparseCore and TensorCore via custom Pallas kernels at terabyte scale.

\para{Join Algorithms on Accelerators}
Efficiently executing joins---the most computationally expensive relational operator---has driven decades of research into hardware-conscious algorithm design. Traditional optimizations focus heavily on Hash Joins~\cite{HardwareOblivious, Sioulas2019}, optimizing hash table bucket layouts and probing strategies to maximize cache hit rates on multi-core CPUs and GPUs. Advanced iterations have explored leveraging fast GPU interconnects (e.g., NVLink) to scale radix partitioning for joins beyond device memory~\cite{Lutz2020, Lutz2022}. However, these techniques rely implicitly on dynamic memory allocation and fine-grained scatter/gather operations. On a TPU, formulating a sparse hash table with empty padding buckets induces severe sparsity, while erratic hash-probing causes DMA queue contention at the memory banks. To circumvent these hardware limitations, \opSortedHashLookupJoin{} abandons sparse hash materialization. Inspired by sort-merge architectures for highly parallel vector engines, it transforms sparse, memory-bound hash probing into packed-key hashing, parallel sorting, and vectorized search.

\para{Other TPU Applications}
While TPUs were designed to accelerate deep learning (e.g., matrix multiplications and embedding lookups~\cite{TPUv4}), recent work has also used them for scalable nearest-neighbor search~\cite{chern2022tpu}. This workload still maps to predictable, iterative tensor computations.
Analytical database queries present a uniquely challenging environment for TPUs due to their data-dependent dynamism, where output sizes and memory access patterns are dictated by the underlying data distribution/values at runtime. To the best of our knowledge, our engine is the first to successfully challenge these static dynamic restrictions, enabling TPUs to execute strict relational algebra effectively across multi-terabyte datasets up to SF 1000.

\section{Conclusion}
Google's TPUs bring about massive computing at scale, yet they have not been sufficiently explored for analytical database workloads.
Our Native TPU Accelerator (NTA) demonstrates immense potential for analytical query processing through HW-conscious algorithm design and JAX/XLA optimizations that leverage SparseCore to efficiently offload dynamic memory accesses, multi-chip TPU slices, and fast ICI links. We show that TPUs have powerful programmable software stacks that enable per-query engineering and unlock effective TPU performance optimizations. Our NTA achieves very competitive TPU-accelerated analytical query processing results (e.g., state-of-the-art performance on TPC-H benchmarks).
With AI/ML workloads consuming all the latest and greatest accelerators, older generation TPUs, like TPU v5p, are naturally becoming more available. It is becoming sensible to migrate database workloads to the earlier-generation accelerators, as they provide massive speedups  for analytical query processing.

As LLMs and semantic AI queries are increasingly integrated into data analytics engines to enable richer structured and unstructured data analytics, migrating SQL workloads to the accelerator opens an unprecedented opportunity. By executing both relational queries and AI semantic operations on a shared computing substrate, our study shows that it is possible to unify enterprise data management with ML/AI accelerator HW/SW stacks, unlocking fundamentally greater performance and efficiency.

\begin{acks}
We would like to thank Chris Taylor, David E. Culler, Greg Ganger, Alon Halevy, 
Cliff Young, Suvinay Subramanian for invaluable feedback and support.
\end{acks}

\bibliographystyle{ACM-Reference-Format}
\bibliography{main}

\end{document}